\documentclass[conference]{IEEEtran}
\IEEEoverridecommandlockouts
\usepackage{cite}
\usepackage{amsmath,amssymb,amsfonts}
\usepackage{algorithmic}
\usepackage{graphicx}
\usepackage{textcomp}
\usepackage{xcolor}
\usepackage{glossaries}
\usepackage{gensymb}
\usepackage{comment}
\usepackage{booktabs}
\usepackage{caption}
\usepackage{subcaption}
\usepackage[normalem]{ulem}
\def\BibTeX{{\rm B\kern-.05em{\sc i\kern-.025em b}\kern-.08em
    T\kern-.1667em\lower.7ex\hbox{E}\kern-.125emX}}
\begin{document}

\title{WILD: a new in-the-Wild Image Linkage Dataset for synthetic image attribution
}

\newacronym{sfhq_t2i}{SFHQ-T2I}{Synthetic Faces High Quality - Text2Image}
\newacronym{ddpm}{DDPM}{Denoising Diffusion Probabilistic Models}
\newacronym{cnn}{CNN}{Convolutional Neural Network}


\begin{centering}
\author{Pietro Bongini$^{\star}$,
Sara Mandelli$^{\dagger}$,
Andrea Montibeller$^{\ddagger}$,
Mirko Casu$^{\mathsection}$,
Orazio Pontorno$^{\mathsection}$,
Claudio Vittorio Ragaglia$^{\mathsection}$\\
Luca Zanchetta$^{\mathparagraph}$,
Mattia Aquilina$^{\mathparagraph}$,
Taiba Majid Wani$^{\mathparagraph}$,
Luca Guarnera$^{\mathsection}$,
Benedetta Tondi$^{\star}$\\
Giulia Boato$^{\ddagger}$,
Paolo Bestagini$^{\dagger}$,
Irene Amerini$^{\mathparagraph}$,
Francesco De Natale$^{\ddagger}$,
Sebastiano Battiato$^{\mathsection}$,
Mauro Barni$^{\star}$\\
\vspace{-7pt}
\and
$^{\star}$ \textit{University of Siena, Department of Information Engineering an Mathematics, Italy}\\
$^{\dagger}$ \textit{Politecnico di Milano, Department of Electronics, Informatics, and Bioengineering, Italy}\\
$^{\ddagger}$ \textit{University of Trento, Department of Information Engineering and Computer Science, Italy}\\
$^{\mathsection}$ \textit{University of Catania, Department of Mathematics and Computer Science, Italy}\\
$^{\mathparagraph}$ \textit{Sapienza University of Rome - Departement of Computer, Control and Management
Engineering, Italy}
}
\end{centering}


\newcommand{\BT}[1]{\textcolor{blue}{{Benedetta: {#1}}}}
\newcommand{\AM}[1]{\textcolor{orange}{{Andrea: {#1}}}}
\newcommand{\MB}[1]{\textcolor{brown}{{Mauro: {#1}}}}

\maketitle

\begin{figure*} 
    \centering
     \subfloat[ADF\label{1a}]{%
       \includegraphics[width=0.098\linewidth]{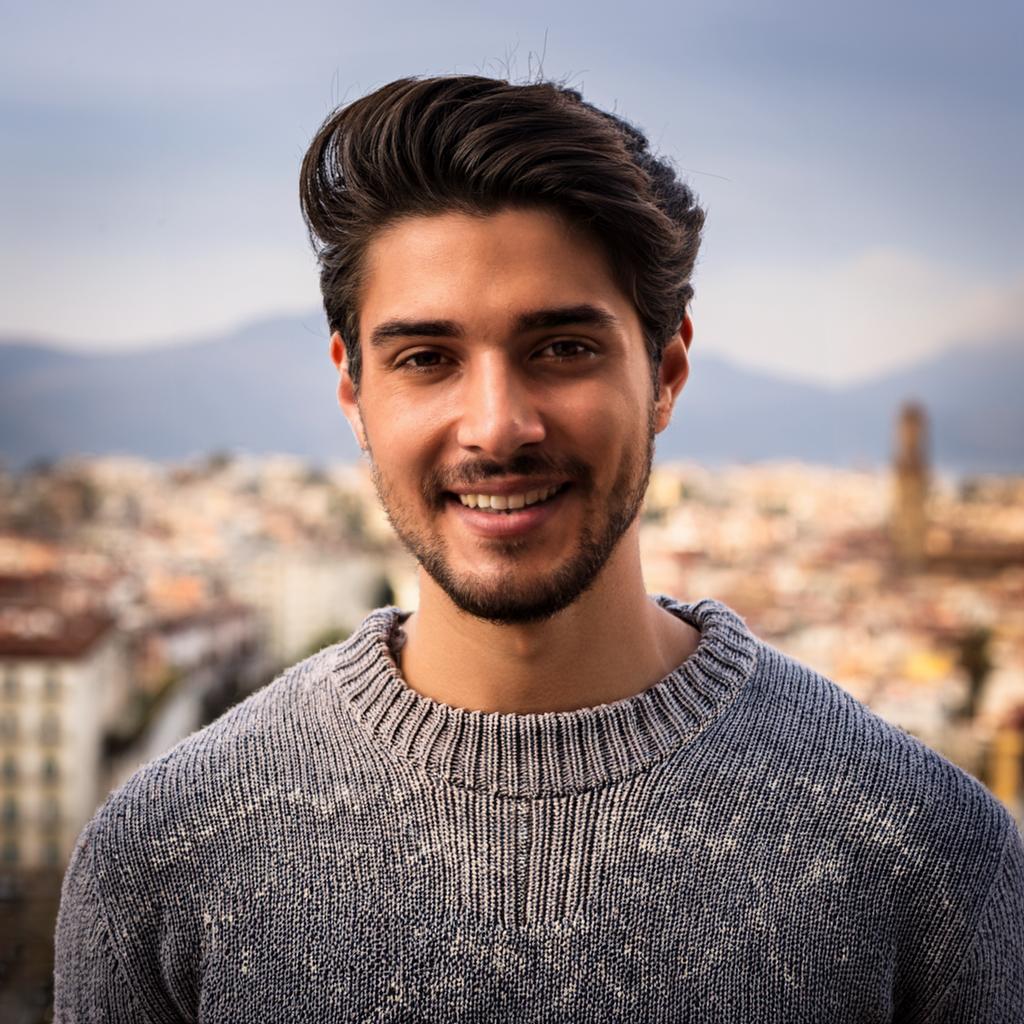}}
    \hfill
    \subfloat[DE3\label{1b}]{%
        \includegraphics[width=0.098\linewidth]{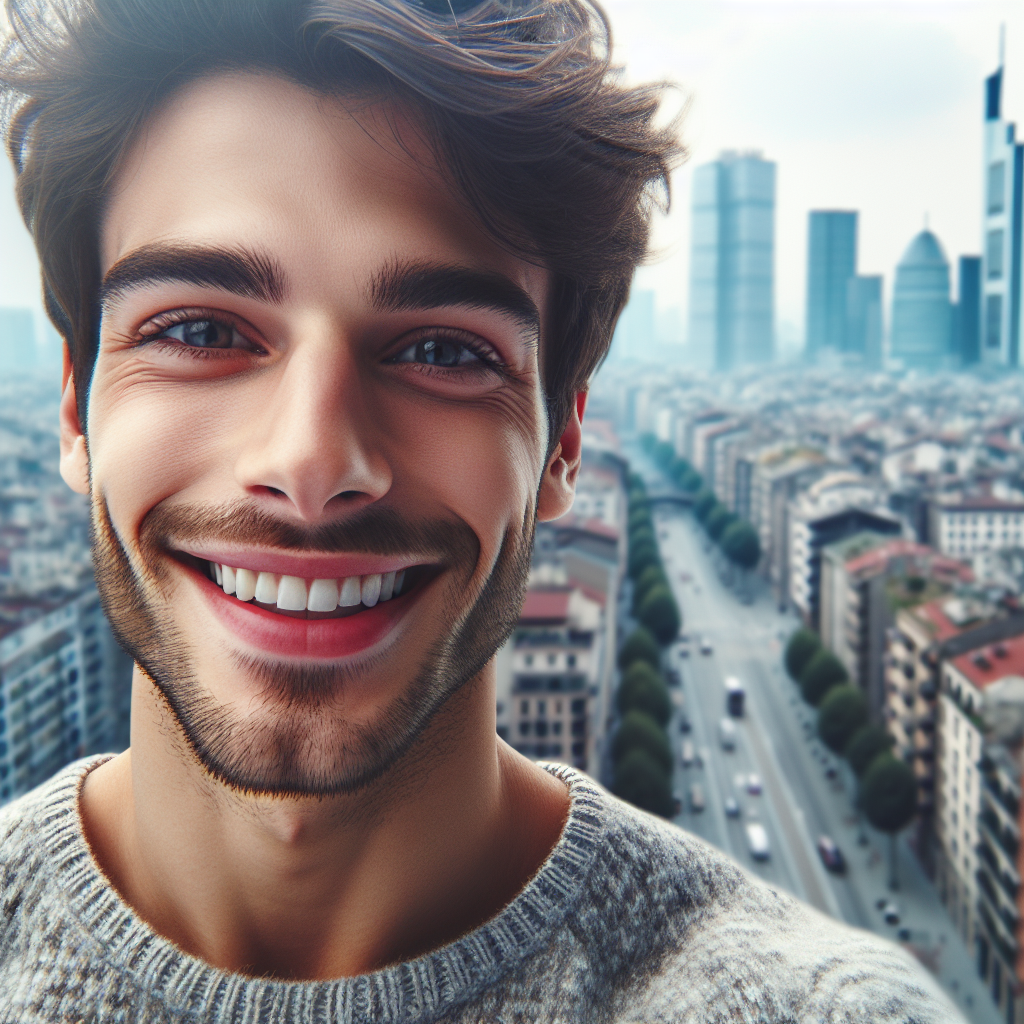}}
    \hfill
    \subfloat[FX1\label{1c}]{%
       \includegraphics[width=0.098\linewidth]{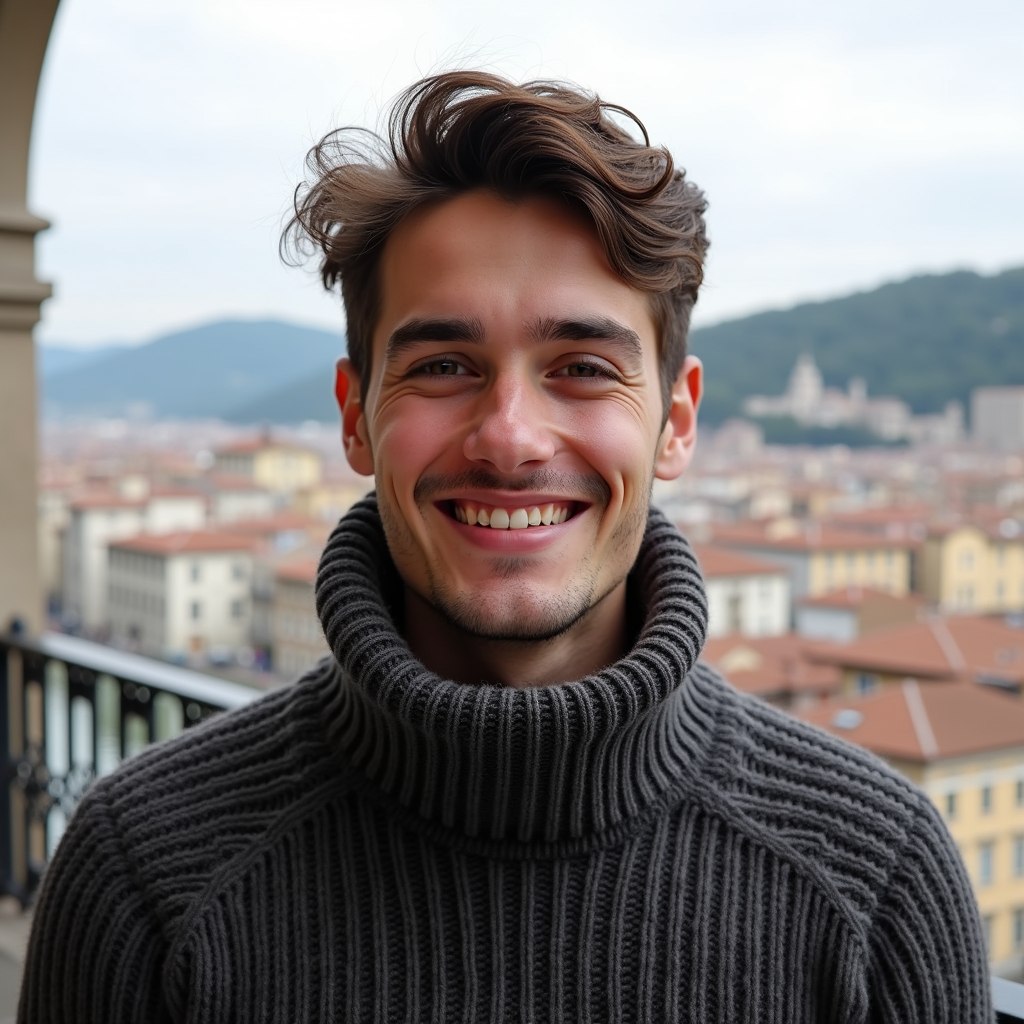}}
    \hfill
    \subfloat[FXP\label{1d}]{%
        \includegraphics[width=0.098\linewidth]{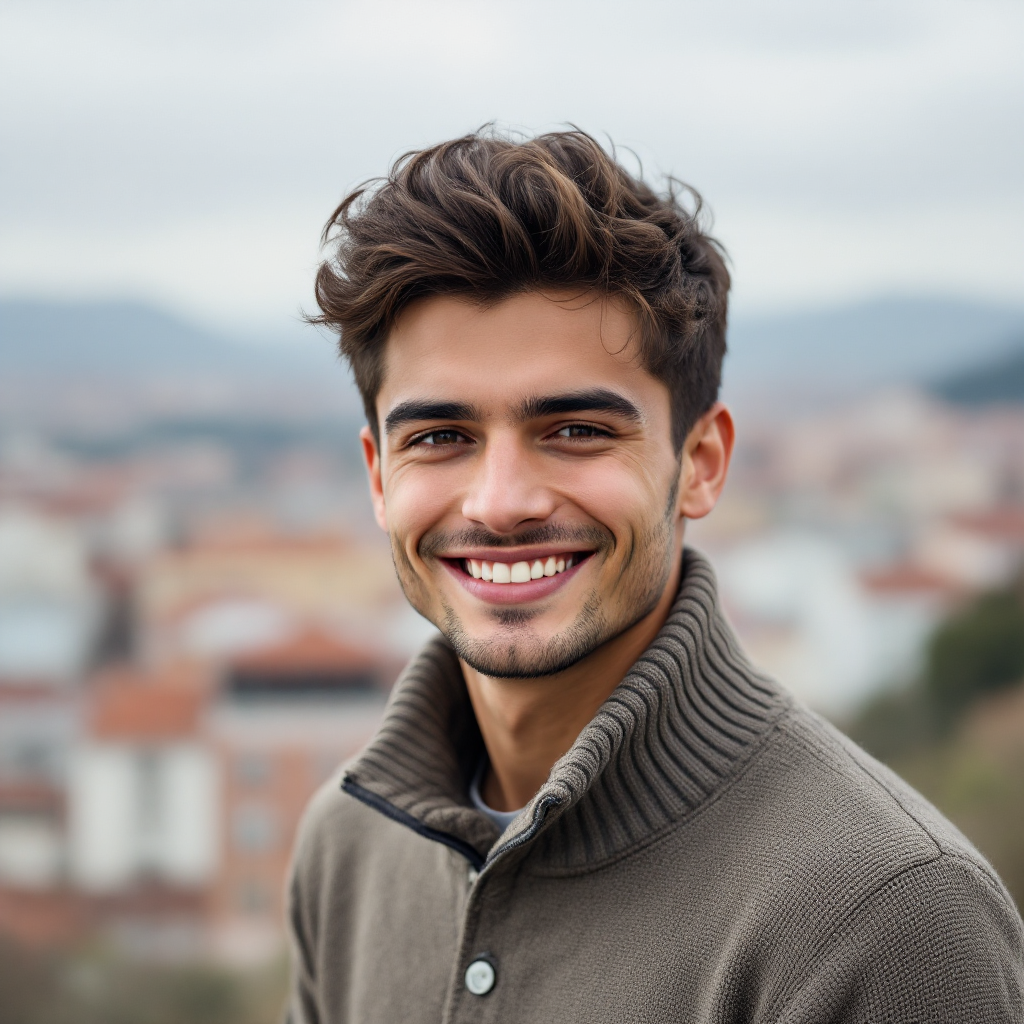}}
    \hfill
    \subfloat[FPK\label{1e}]{%
        \includegraphics[width=0.098\linewidth]{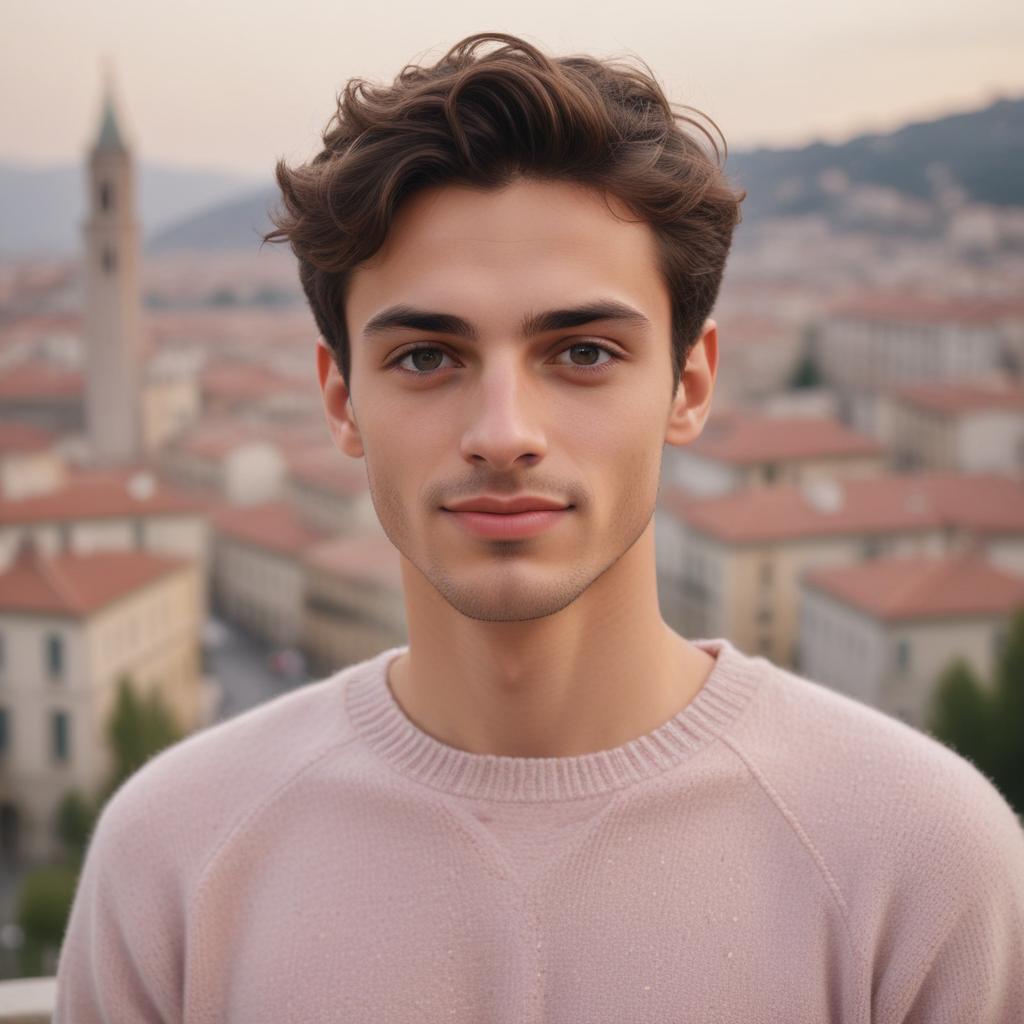}}
    \hfill
    \subfloat[LEO\label{1f}]{%
        \includegraphics[width=0.098\linewidth]{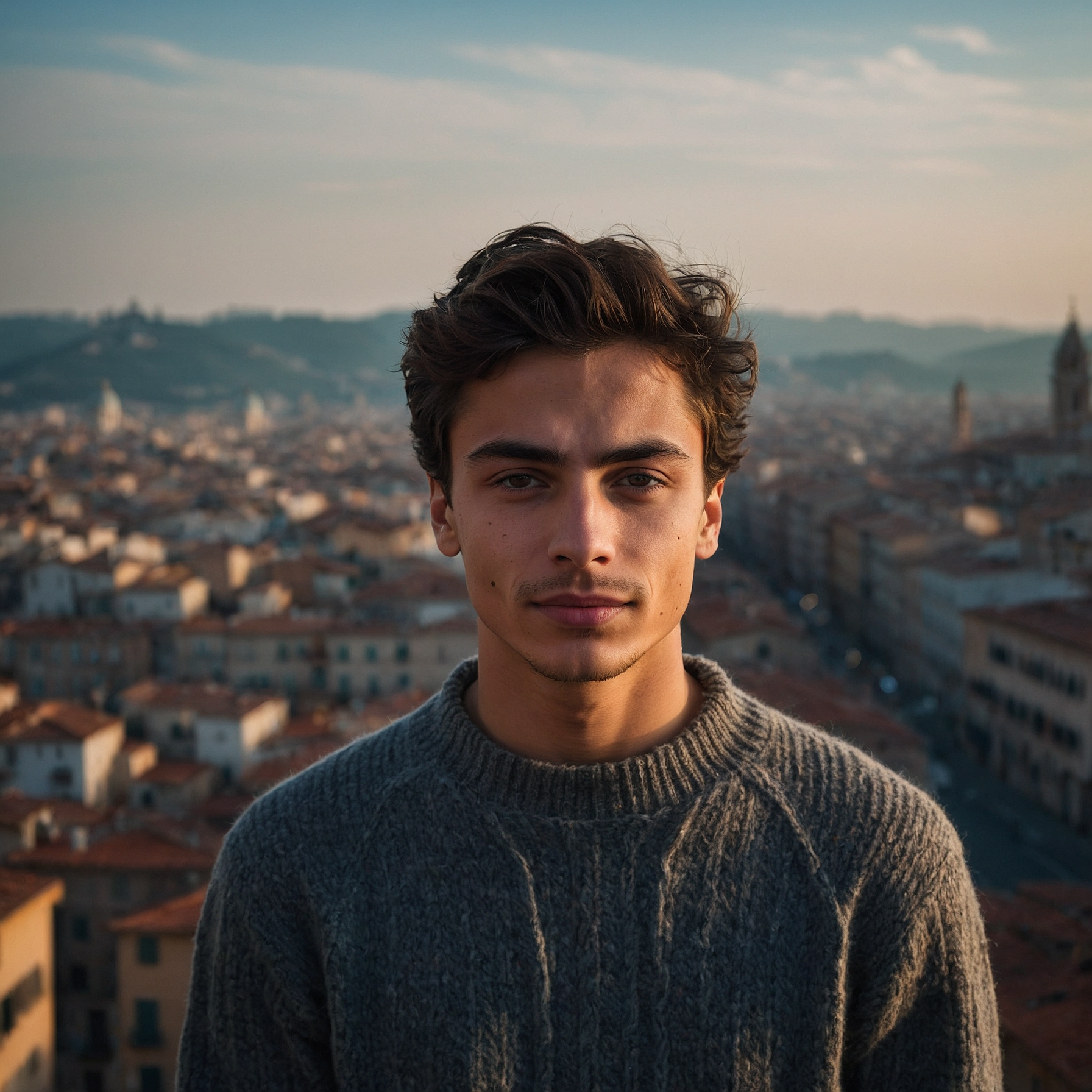}}
    \hfill
    \subfloat[MDJ\label{1g}]{%
        \includegraphics[width=0.098\linewidth]{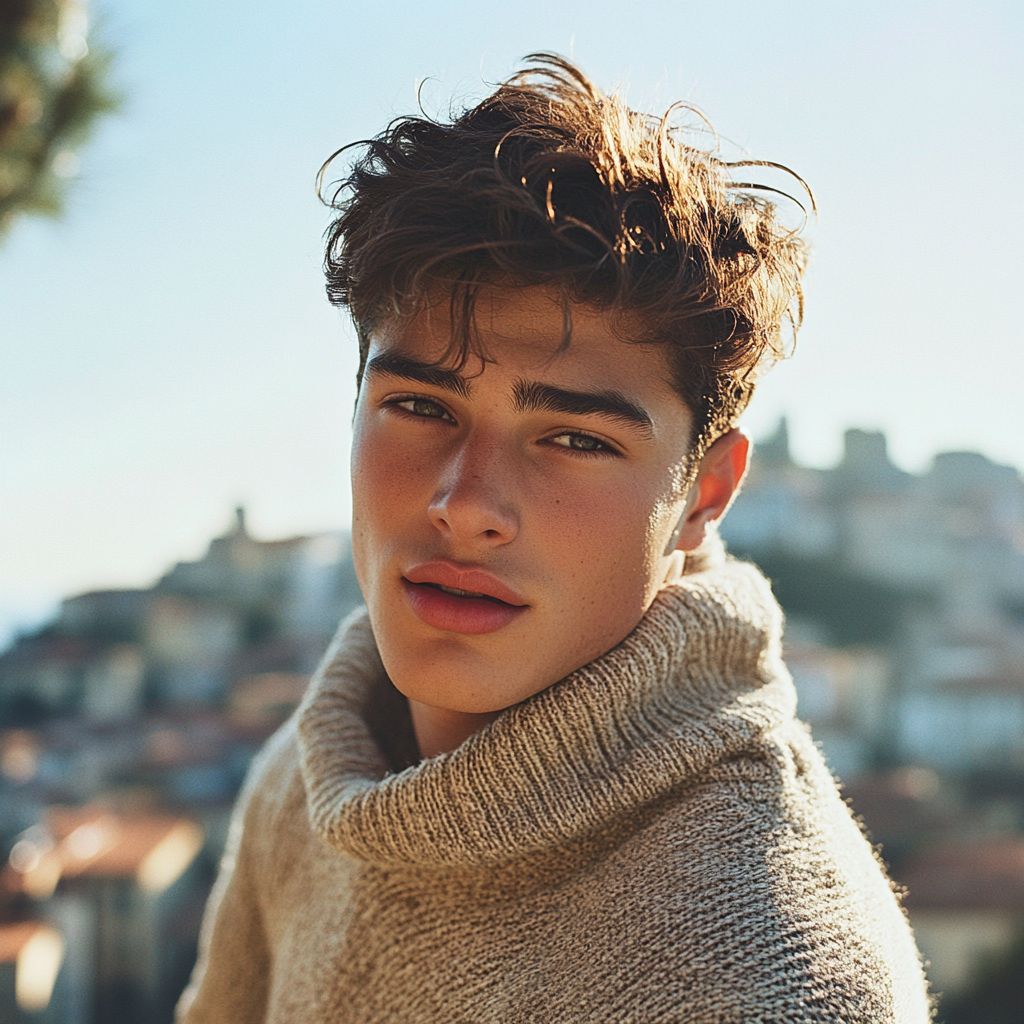}}
    \hfill
    \subfloat[S35\label{1h}]{%
       \includegraphics[width=0.098\linewidth]{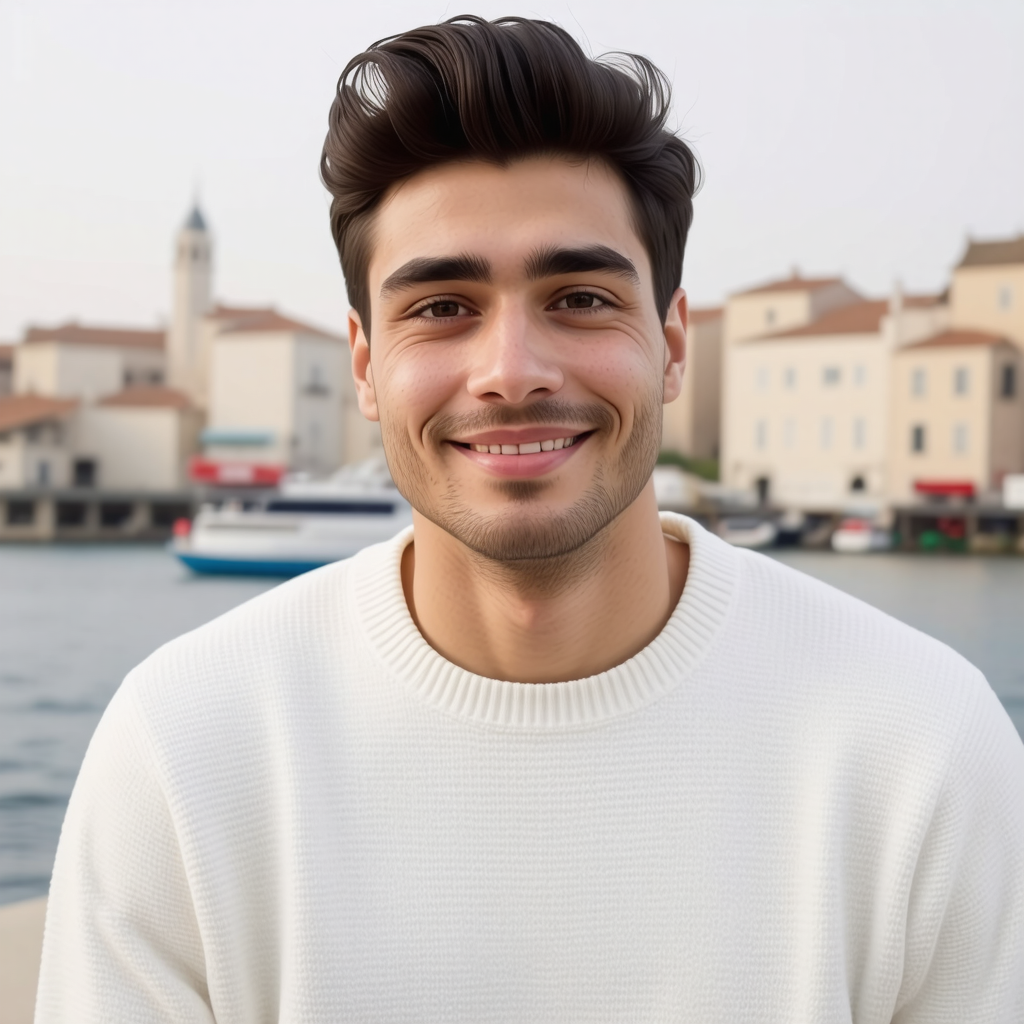}}
    \hfill
    \subfloat[SXL\label{1i}]{%
        \includegraphics[width=0.098\linewidth]{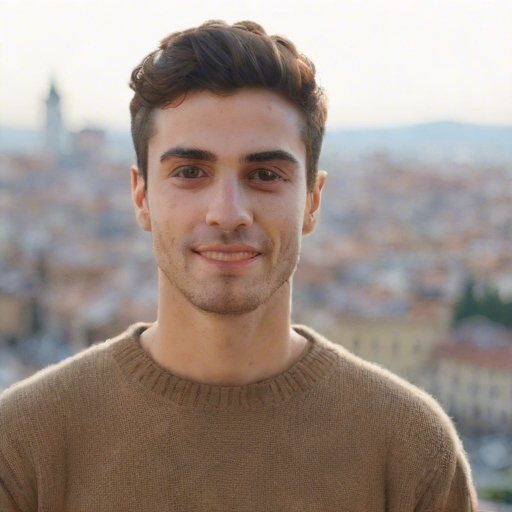}}
    \hfill
    \subfloat[STR\label{1j}]{%
        \includegraphics[width=0.098\linewidth]{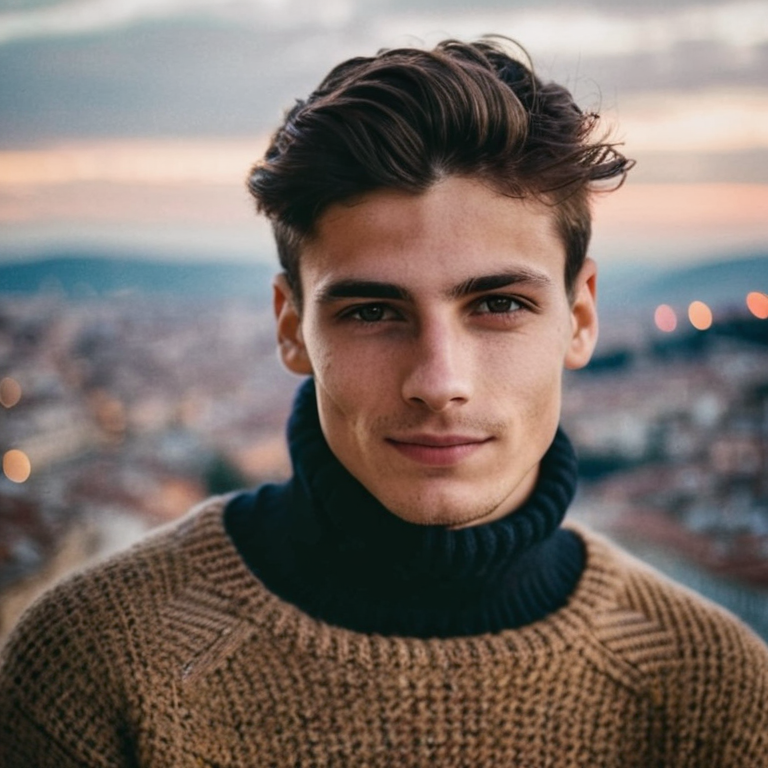}}
    \\
    \subfloat[ADF\label{2a}]{%
       \includegraphics[width=0.098\linewidth]{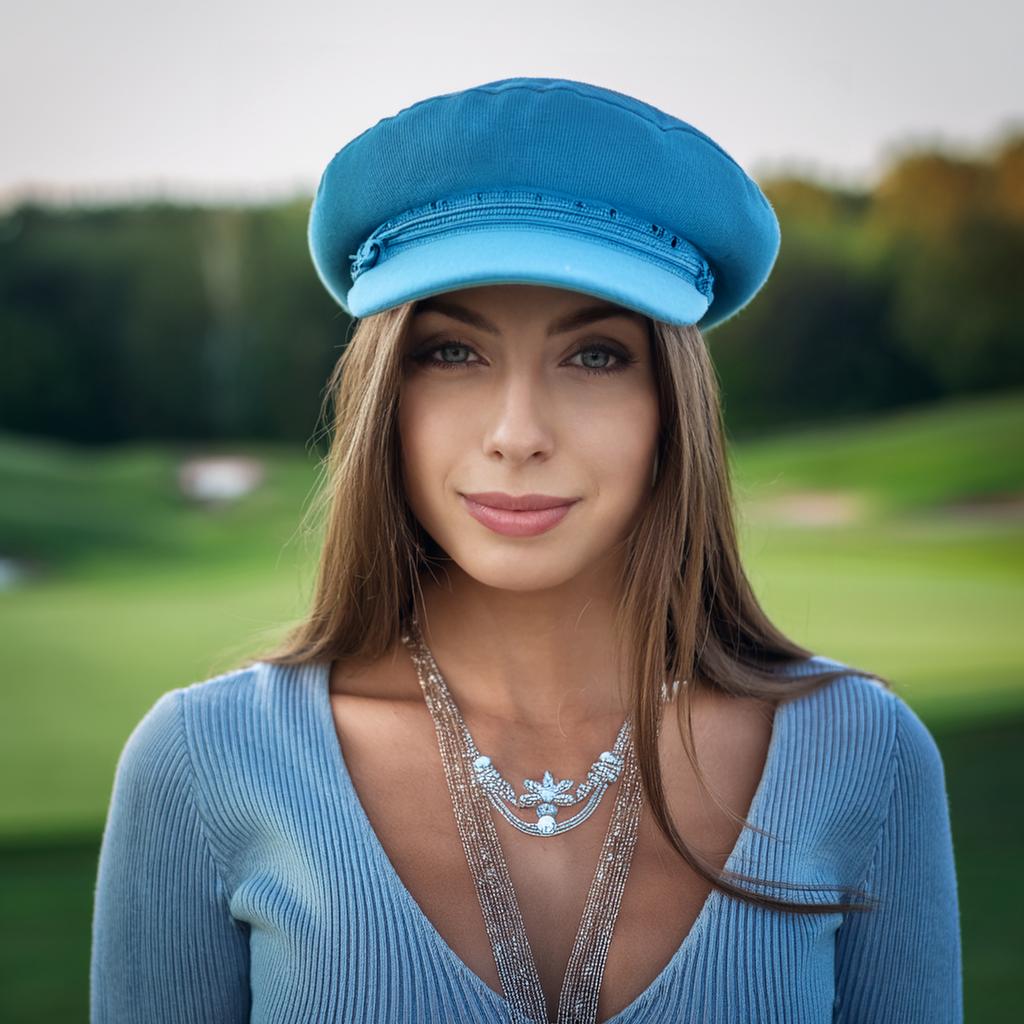}}
    \hfill
    \subfloat[DE3\label{2b}]{%
        \includegraphics[width=0.098\linewidth]{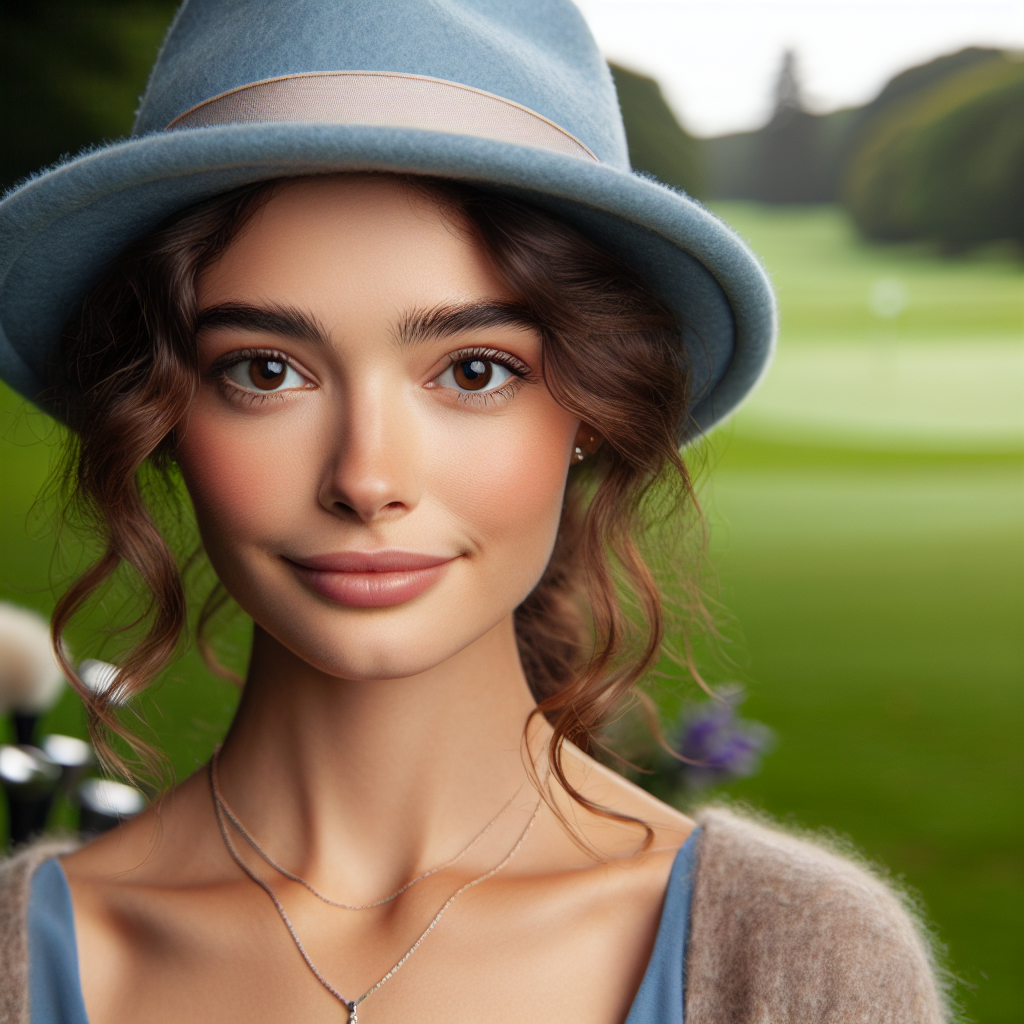}}
    \hfill
    \subfloat[FX1\label{2c}]{%
       \includegraphics[width=0.098\linewidth]{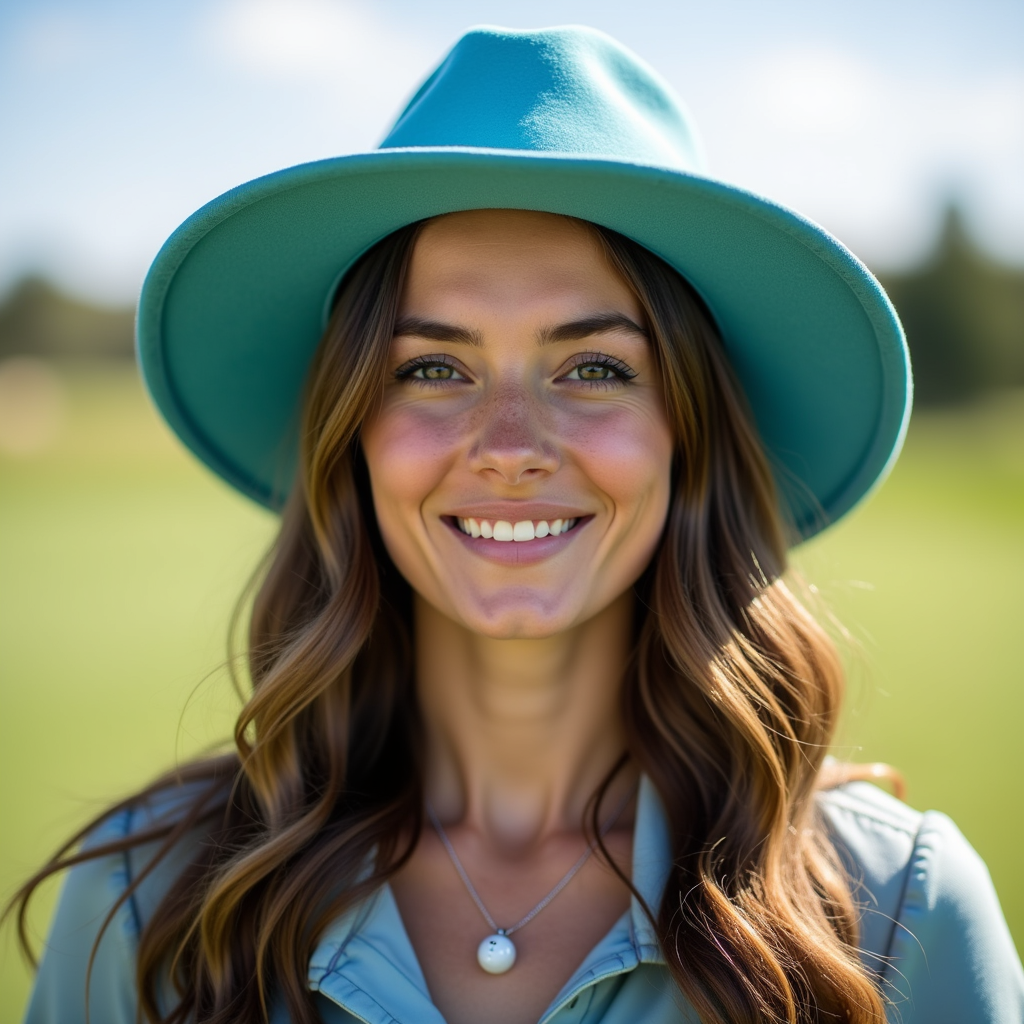}}
    \hfill
    \subfloat[FXP\label{2d}]{%
        \includegraphics[width=0.098\linewidth]{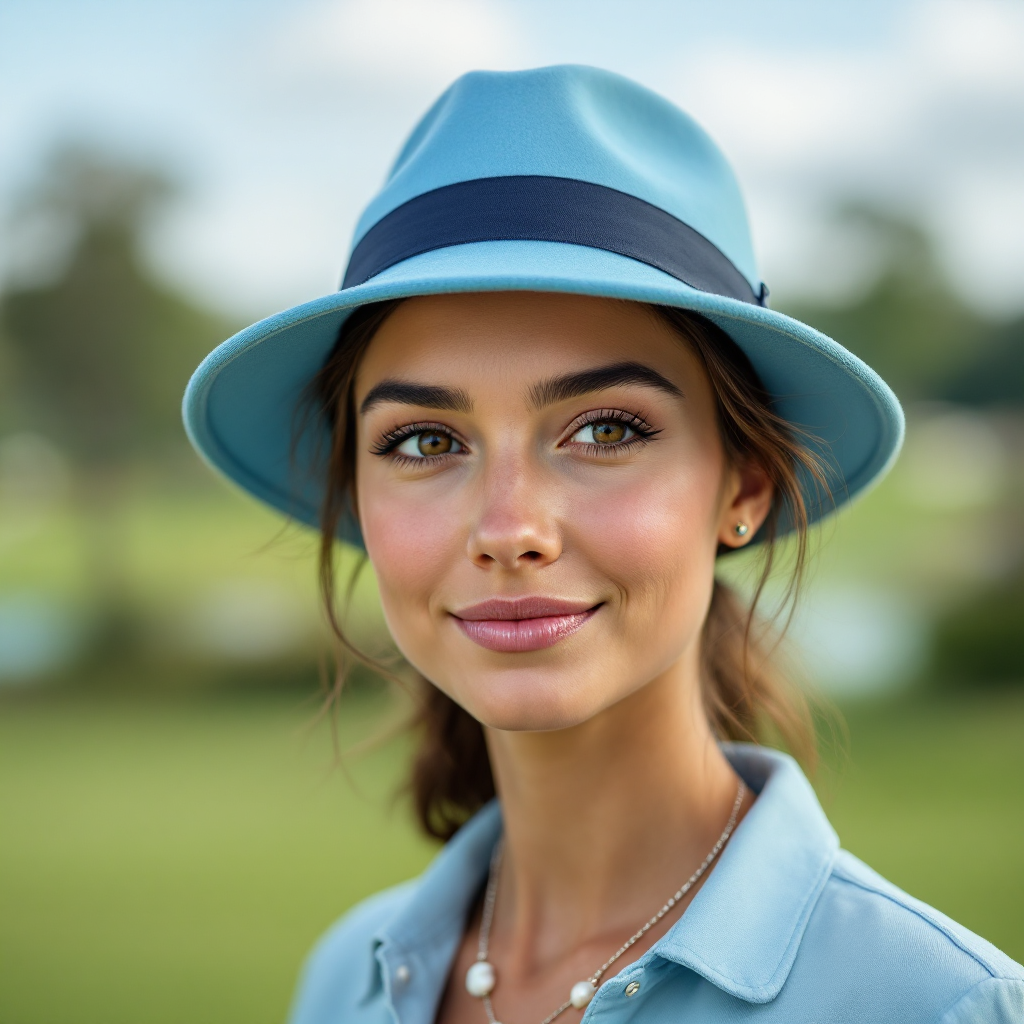}}
    \hfill
    \subfloat[FPK\label{2e}]{%
        \includegraphics[width=0.098\linewidth]{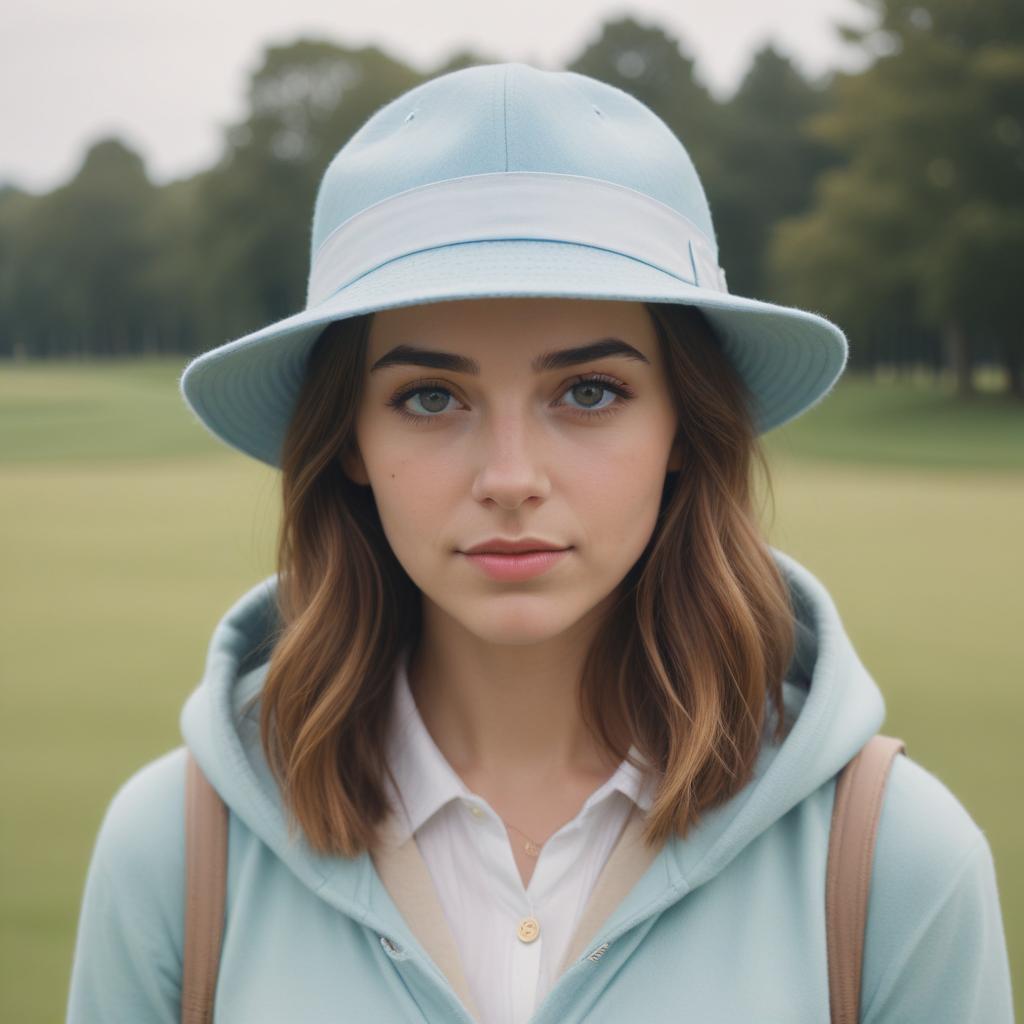}}
    \hfill
    \subfloat[LEO\label{2f}]{%
        \includegraphics[width=0.098\linewidth]{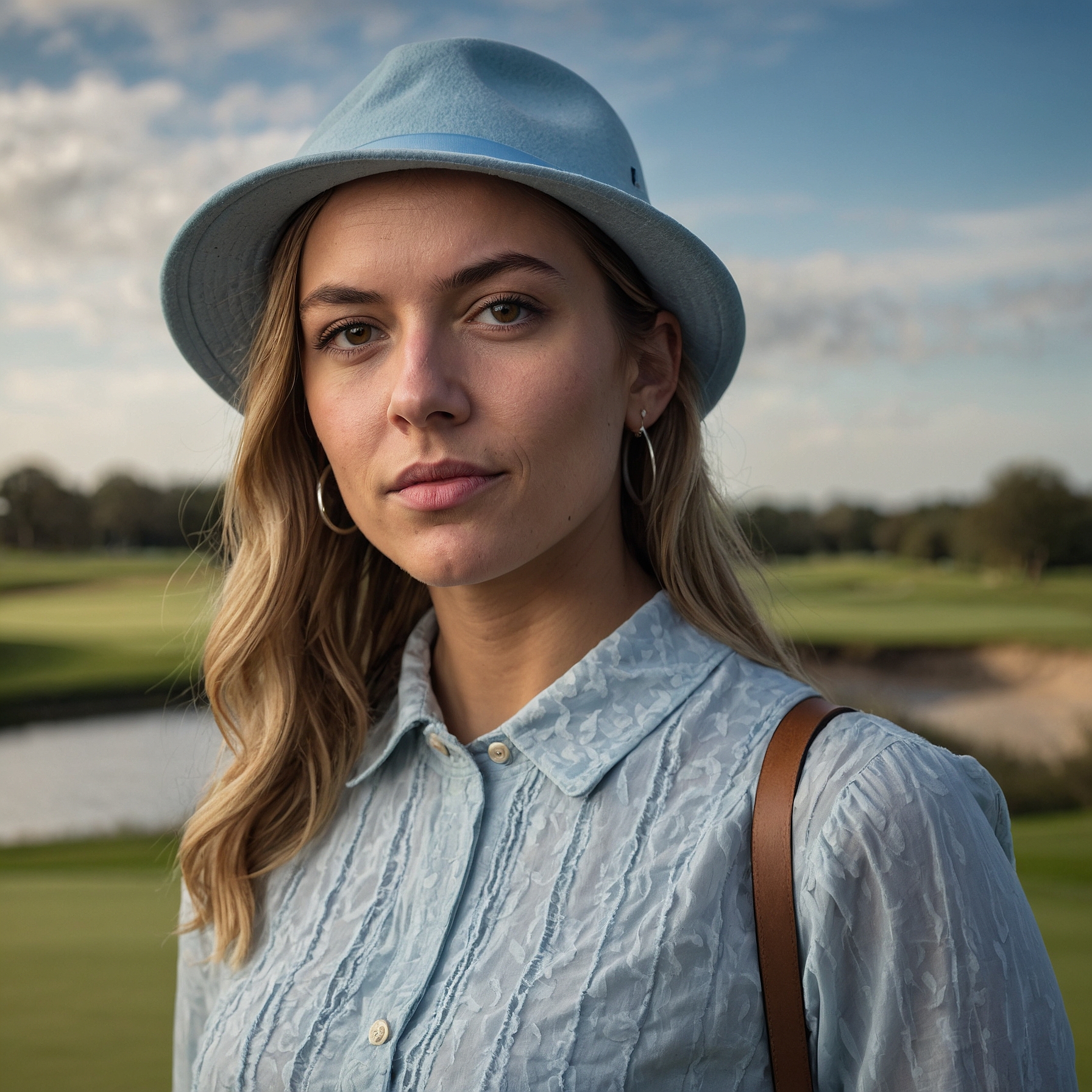}}
    \hfill
    \subfloat[MDJ\label{2g}]{%
        \includegraphics[width=0.098\linewidth]{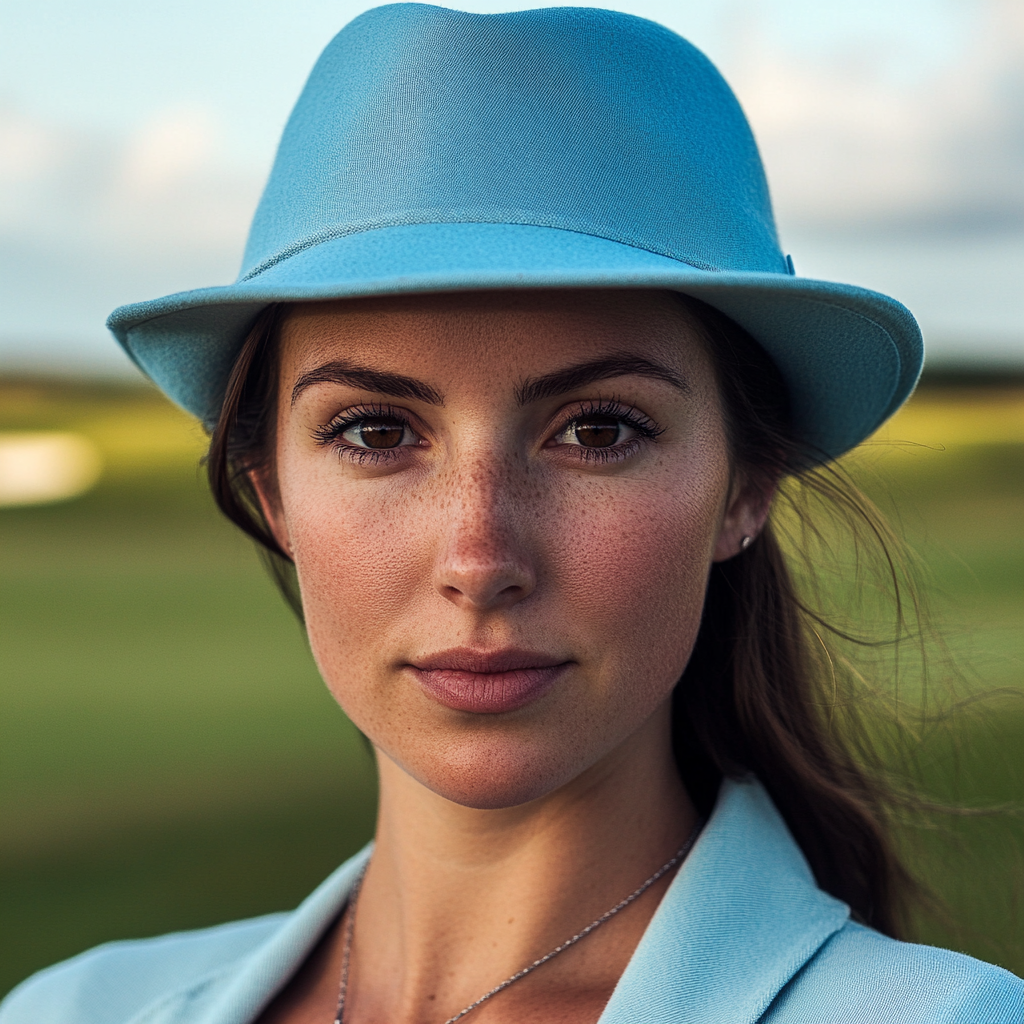}}
    \hfill
    \subfloat[S35\label{2h}]{%
       \includegraphics[width=0.098\linewidth]{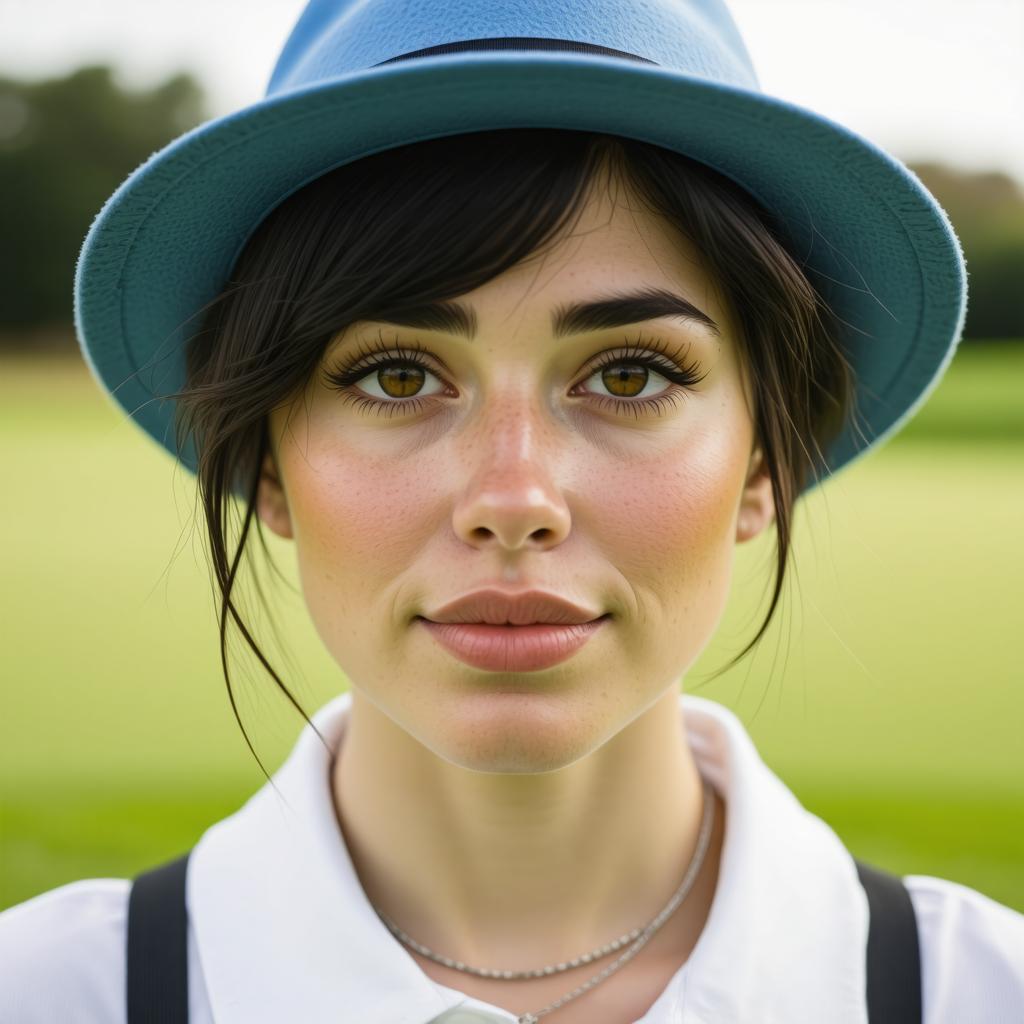}}
    \hfill
    \subfloat[SXL\label{2i}]{%
        \includegraphics[width=0.098\linewidth]{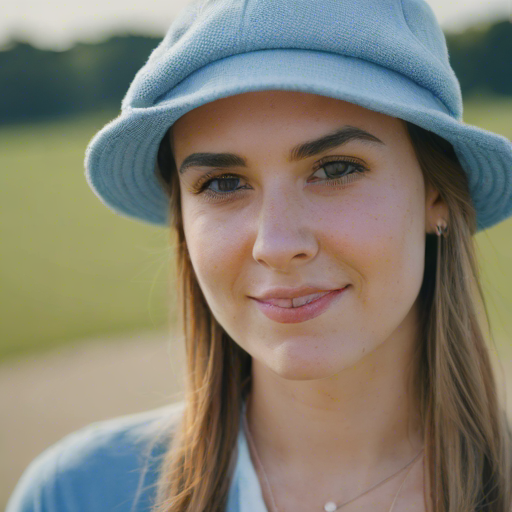}}
    \hfill
    \subfloat[STR\label{2j}]{%
        \includegraphics[width=0.098\linewidth]{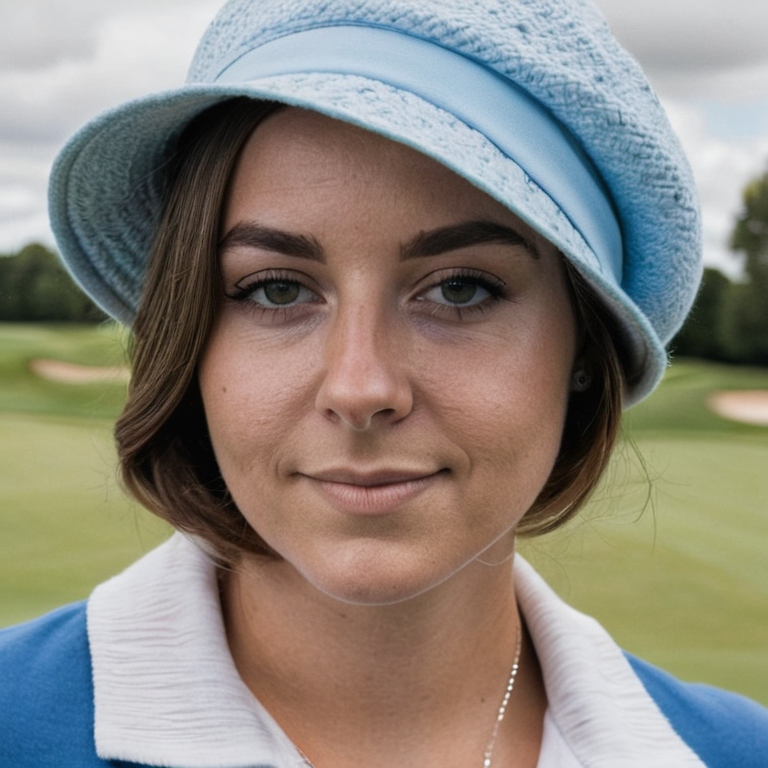}}
    \\
  \caption{Two groups of ten images generated by the closed set generators with the same text prompt. The generators are identified by the codes described in Section~\ref{subsec:generators}. The prompts used to generate the images are: (top, a-j): prompt 439, ``Image of an Italian young man who is wearing a pullover. He has a pointy face and full lips. The subject is looking into the camera in happiness. The image is taken with a city in the background''; (bottom, k-t): prompt 266, “A proud British woman in her 20s with hooded brown eyes and a small nose is looking into the camera. She is wearing a sky blue fedora and a necklace. The subject is portrayed with a golf course in the background". }
  \label{fig:prompt_example} 
  \vspace{-10pt}
\end{figure*}

\begin{abstract}
Synthetic image source attribution is an open challenge, with an increasing number of image generators being released yearly. The complexity and the sheer number of available generative techniques, as well as the scarcity of high-quality open source datasets of diverse nature for this task, make training and benchmarking synthetic image source attribution models very challenging. WILD\footnote{The dataset is available at: https://www.kaggle.com/datasets/pietrob92/wild-in-the-wild-image-linkage-dataset} is a new in-the-Wild Image Linkage Dataset designed to provide a powerful training and benchmarking tool for synthetic image attribution models. The dataset is built out of a closed set of 10 popular commercial generators, which constitutes the training base of attribution models, and an open set of 10 additional generators, simulating a real-world in-the-wild scenario. Each generator is represented by 1,000 images, for a total of 10,000 images in the closed set and 10,000 images in the open set. Half of the images are post-processed with a wide range of operators. WILD allows benchmarking attribution models in a wide range of tasks, including closed and open set identification and verification, and robust attribution with respect to post-processing and adversarial attacks. Models trained on WILD are expected to benefit from the challenging scenario represented by the dataset itself. Moreover, an assessment of seven baseline methodologies on closed and open set attribution is presented, including robustness tests with respect to post-processing.
\end{abstract}


\vspace{-2pt}
\section{Introduction}
\label{sec:intro}
Advancements in Generative AI (GenAI), particularly Diffusion Models (DMs) and Generative Adversarial Networks (GANs), have revolutionized multimedia content creation, producing highly realistic synthetic images that blur the line with reality~\cite{casu2024genai, Corvi2023Intriguing, Hirte2021Realistic}. 
These innovations, while impressive, challenge multimedia forensics by complicating the attribution of synthetic images to their source generators, a task that is both intricate and essential.
%
In fact, synthetic image attribution (sometimes referred to as image linkage), that is, the identification of the generative model used to produce a given image, is vital for forensic analysis, intellectual property protection, and counter AI-driven misinformation and impersonation.
However, current methods struggle to generalize, especially in open-set scenarios involving unknown or new generators\cite{khoo2022deepfake,bindini2024tiny}. 

Some datasets are available that can be used to address this task.
For example, the \gls{sfhq_t2i} dataset~\cite{sfhq_t2i} provides ~$120,000$ high-quality synthetic faces ($1024 \times 1024$ resolution) from models such as FLUX.1 (Pro, dev, schnell), Stable Diffusion XL, and DALL$\cdot$E\,3.
However, this dataset presents a strongly uneven image distribution across generators and lacks prompt-image linkage to mirror real-world attribution difficulties.
The ArtiFact dataset, proposed in~\cite{artifact}, comprises $25$ diverse generators, object categories, and real-world challenges. Nonetheless, ArtiFact's image resolution is very limited (i.e. $200 \times 200$) and, once again, images are not uniformly distributed among generators. 
The GenImage dataset~\cite{zhu2023genimage} comprises eight different generative models, each used to produce synthetic images for the $1,000$ distinct labels in ImageNet~\cite{deng2009imagenet}, ensuring a nearly equal distribution across classes. In contrast, this dataset does not include any form of image processing, resembling only ideal laboratory conditions.
Moreover, none of these datasets include an open set for effective simulations of what attribution models trained on them can achieve in the wild.
Additionally, commercial generators, even the most popular and utilized ones, are scarce if present at all in such datasets: among the datasets cited above, the only examples are Midjourney in GenImage and DALL$\cdot$E\,3 in \gls{sfhq_t2i}.

The lack of datasets containing high-resolution images, uniformly distributed across synthetic generators (including commercial ones as well), with editing operations applied, and of an open set for simulating in-the-wild conditions, 
prevents the possibility of evaluating attribution models on realistic scenarios.
To tackle these issues, we present WILD, a novel “in-the-Wild" Image Linkage Dataset tailored for synthetic image attribution. The dataset focuses on the source attribution, without taking into account real vs fake detection.
The dataset is composed exclusively of head-and-shoulder images of synthetic people from fixed prompts. WILD features a closed set containing images produced by $10$ popular text-to-image generators ($1,000$ images each), among which six commercial platforms, and an open set section, obtained by using $10$ other generators contributing $1,000$ images each. 

In the closed set, image linkage allows to retrace the prompt each image was produced with. This is achieved by generating $1,000$ prompts with an ad-hoc prompting script and using each prompt once on every generator. 
The open set is instead conceived as a simulation of what can happen in the wild. It is composed of generators with very different characteristics, in order to better cover the possible image features that can be encountered by the attribution models once in use. 
To reflect real-world conditions, half of the images have undergone three post-processing steps (e.g., rotation, resizing, compression, and many others): for these images, we release the plain version, and the post-processed versions with one, two, and three operators.
The contributions of this paper are:
\begin{itemize}
\item We release WILD, a challenging source attribution dataset of head-and-shoulder synthetic faces. 
It includes a closed set produced by some of the most popular synthetic generators. The images are generated through randomized and evenly distributed prompts across generators. This linkage between images and prompts helps to minimize biases and improve the understanding of both generation and attribution mechanisms. WILD also includes a large and varied open set of synthetic human faces, simulating realistic in-the-wild scenarios.
\item To better reflect images from the real-world, e.g., those shared through social media platforms, we apply several common post-processing operations. In this way, we allow testing the robustness of forensic detectors. 
\item We present the results of several experiments obtained by applying state-of-the-art source attribution baselines to the WILD dataset. This allows to assess the dataset characteristics and to benchmark new attribution models based on their results. The experiments include closed-set attribution, open-set attribution and robustness evaluation.
\end{itemize}
\vspace{-5pt}


\section{WILD Dataset}
\label{sec:dataset}
WILD is designed as an Image Linkage dataset for benchmarking source attribution methods, both in closed set and open set scenarios. To simulate real-world conditions, the closed set includes some of the most popular commercial and open-source text-to-image generators. The open set, instead, contains a large and varied collection of different image generators (including both DMs and GANs), to summarize the widest possible range of generative techniques. 
The images of the dataset were generated with a total of $20$ generators, $10$ for the closed set and $10$ for the open set. 
Each generator is represented by $1,000$ images, for a total of $20,000$ images. Half of the images in each subset were post-processed by one, two, or three image-processing operators, for a grand total of $50,000$ images.
In the following, we provide additional details on the composition of the dataset and its construction procedure.

\subsection{Closed Set}
\label{subsec:closed_set}
The closed set was built using $10$ of the most popular commercial and open source text-to-image generators. Each generator was used to generate $1,000$ uncompressed images. 
To avoid significant biases, we developed an ad-hoc prompting methodology.
Each prompt was used to generate $10$ images, one for each generator. In this way, we ensured that every class contains the same data distributions. In addition, in this way we created a direct link between each image and its prompt.
In Fig.~\ref{fig:prompt_example}, we report some examples of images generated with the same text prompt from each generator. 

\subsection{Closed set generators}
\label{subsec:generators}
The generators used to build the closed set are listed below, together with a description of the hyperparameters used. For the sake of readability, in the tables and figures of the following sections, these models are indicated by the three-character codes introduced here.
\subsubsection{Adobe Firefly (ADF)}
\label{subsubsec:firefly}
this is the proprietary solution released by Adobe to generate extremely realistic images and edit existing photos using text descriptions~\cite{firefly}. Specifically, we selected the second version of the Firefly text-to-image generator, which allows to create $1024 \times 1024$ images. 
\subsubsection{DALL$\cdot$E\,3 (DE3)}
\label{subsubsec:dalle3}
this is a commercial text-to-image generator by Open-AI, ensuring consistency between the textual description and the visual output~\cite{dalle3}.
We generated $1024 \times 1024$ images with the standard quality settings.

\subsubsection{FLUX.1 (FX1)}
\label{subsubsec:flux.1}
this is a diffusion-based text-to-image framework freely provided by Black Forest Labs, optimized for high-resolution, photorealistic outputs with minimal computational cost~\cite{flux2024}. 
For dataset creation, we generated $1024\times1024$ images using a guidance scale of $3.5$, $30$ inference steps, and a $512$-token max sequence length.
\subsubsection{FLUX 1.1 Pro (FXP)}
\label{subsubsec:flux.pro}
this is an advanced diffusion-based text-to-image framework by Black Forest Labs, designed for fast, high-resolution, photorealistic generation ~\cite{fluxpro}. We generated $1024\times1024$ images,
using a fixed seed of $42$, a safety tolerance of $2$, and without prompt upsampling. 
%
\subsubsection{Freepik (FPK)}
\label{subsubsec:Freepik}
this is a commercial text-to-image generative model~\cite{freepik}. We generated $1024\times1024$ images with parameters: $8$ inference steps, a guidance scale of $1.0$, a photo style with pastel color, warm lighting, and portrait mode.

\subsubsection{Leonardo AI (LEO)}
\label{subsubsec:Leonardo}
this is a commercial text-to-image generator~\cite{leonardoai}. The images were produced using the Kino XL model with the following parameters: a size of $1024 \times 1024$, portrait style, a guidance scale of $7$, $15$ inference steps, and the ``alchemy'', ``photoReal'' and ``photoRealVersion'' parameters set to ``True'', ``True'' and ``v2'', respectively.

\subsubsection{Midjourney (MDJ)}
\label{subsubsec:Midjourney}
this is a commercial text-to-image generative framework~\cite{midjourney}.
We used model version $6.1$ and a size of $1024 \times 1024$.
In this case, a negative prompt was specified to avoid excessively unrealistic images\footnote{Negative prompt used for Midjourney: \textit{unrealistic, cartoon, anime, painting, illustration, greyscale, sepia, drawing, sketch, fantasy, elves, orcs, strange nose.}}.
\subsubsection{Stable Diffusion 3.5 Large (S35)}
\label{subsubsec:SD35}
this is an advanced text-to-image model freely provided by StabilityAI~\cite{rombach2022high, stable_diffusion_35_large}. We used $50$ inference steps, a guidance scale of $3.0$, and a $512$-token max sequence length to produce $1024\times1024$ images.
\subsubsection{Stable Diffusion XL Turbo (SXL)}
\label{subsubsec:SDXL}
this is another text-to-image model freely released by StabilityAI~\cite{sdxl_turbo}. 
To ensure a good visual quality for the synthesized images, we used a fixed resolution of $512\times512$ pixels and $2$ inference steps. 
\subsubsection{Starry AI (STR)}
\label{subsubsec:starry}
this is a commercial text-to-image generator~\cite{starryai}. 
We generated $1024\times1024$ images without high-resolution enhancement and using $20$ inference steps.
\subsection{Prompt Generation}
\label{subsec:prompts}
The same $1,000$ head-and-shoulder prompt descriptions were used for all the generators in the closed set.
In this scope, it is expected that an attribution model can rely on semantic image features that might be relative to the person, clothing, background, or maybe on low-level features that persist throughout the whole image. As a consequence, it is fundamental to introduce a certain variability in the images across all these characteristics. This variability was injected through the text prompts.
We generated the prompts automatically. A simple token system allowed various characteristics to be introduced in each prompt, randomizing or rotating them. Five different sentence structures were used to play with the different nuances of interpretation each model can have for different text structures. Prompts were generated in blocks of $100$, rotating every sentence structure twice (for a total of $200$ prompts per structure).
The sentence structures can be summarized as follows\footnote{The parts in round brackets, describing expression and background, were not always included to increase variability. The parts in angle brackets represent characteristics that vary across prompts.}:
\begin{itemize}
\item A: $<$subject$>$ with $<$facial features$>$ wearing $<$clothing$>$ $<$be$>$ looking $<$looking direction$>$ (in $<$expression(emotion)$>$) ($<$background$>$).
\item B: Picture of a/an ($<$expression(adjective)$>$) $<$subject$>$ with $<$facial features$>$. $<$pronoun$>$ $<$wear$>$ $<$clothing$>$ and $<$look$>$ $<$looking direction$>$ ($<$background$>$).
\item C: Image of $<$subject$>$ who is wearing $<$clothing$>$ (, $<$background$>$). The subject has $<$facial features$>$ and (shows $<$expression(emotion)$>$ while)/(is) looking $<$looking direction$>$.
\item D: Head and shoulder picture of ($<$expression(adjective)$>$) $<$subject$>$ with $<$facial features$>$. $<$pronoun$>$ is looking $<$looking direction$>$ ($<$background$>$) and is wearing $<$clothing$>$.
\item E: Portrait of ($<$expression(adjective)$>$) $<$subject$>$ with $<$facial features$>$. $<$pronoun$>$ $<$wear$>$ $<$clothing$>$. The subject $<$look$>$ $<$looking direction$>$ ($<$background$>$).
\end{itemize}

\begin{figure*}[t]
    \centering
     \subfloat[DE1\label{aa}]{%
       \includegraphics[width=0.098\linewidth]{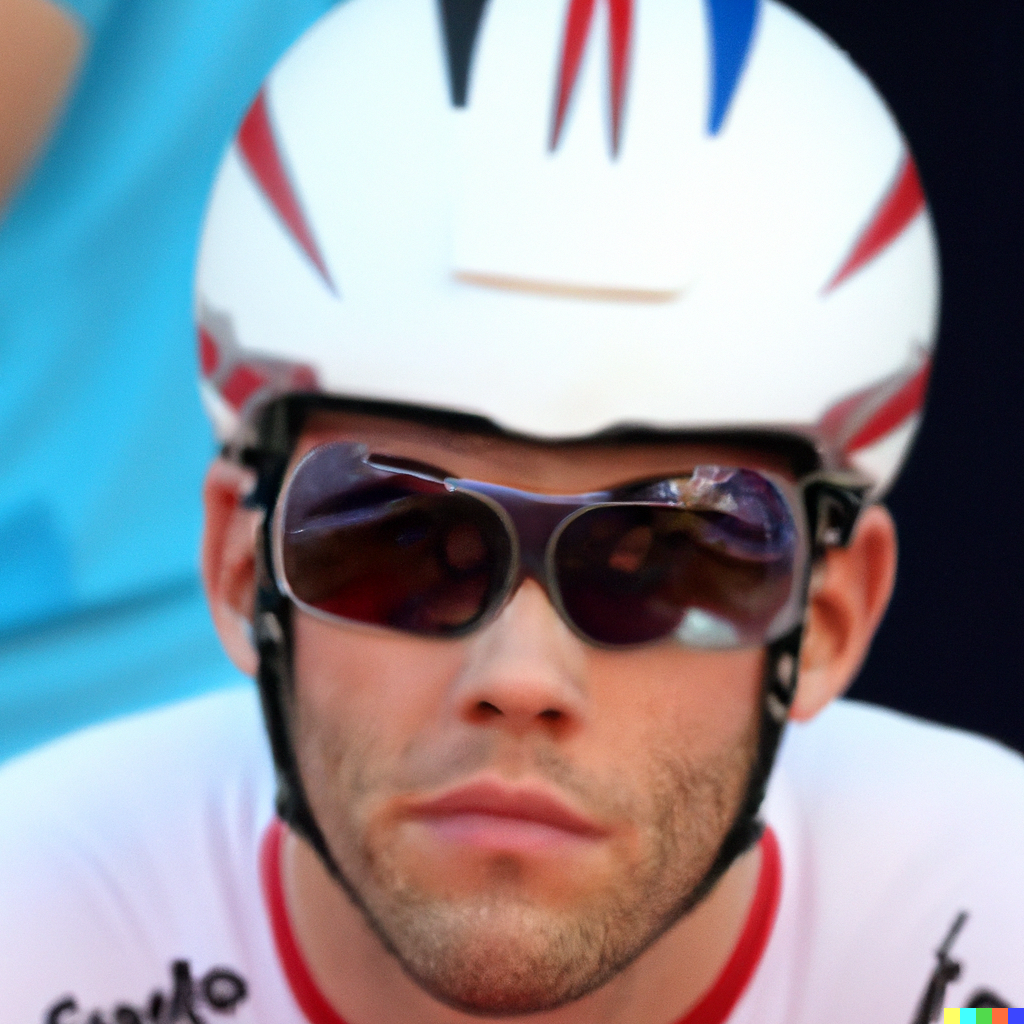}}
    \hfill
    \subfloat[DPA\label{ab}]{%
        \includegraphics[width=0.098\linewidth]{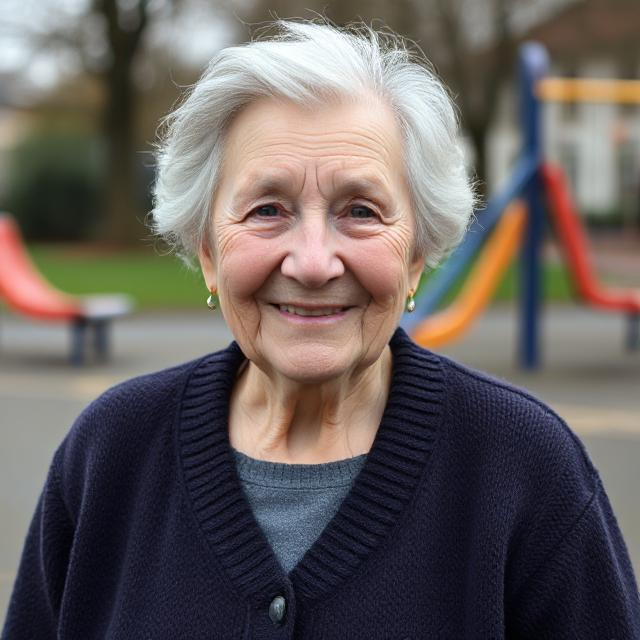}}
    \hfill
    \subfloat[HPA\label{ac}]{%
       \includegraphics[width=0.098\linewidth]{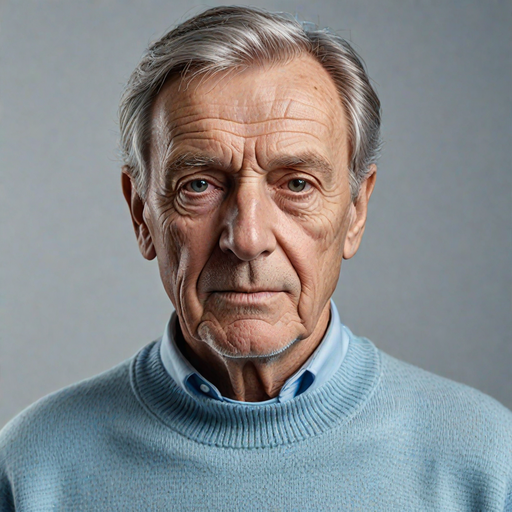}}
    \hfill
    \subfloat[NVS\label{ad}]{%
        \includegraphics[width=0.098\linewidth]{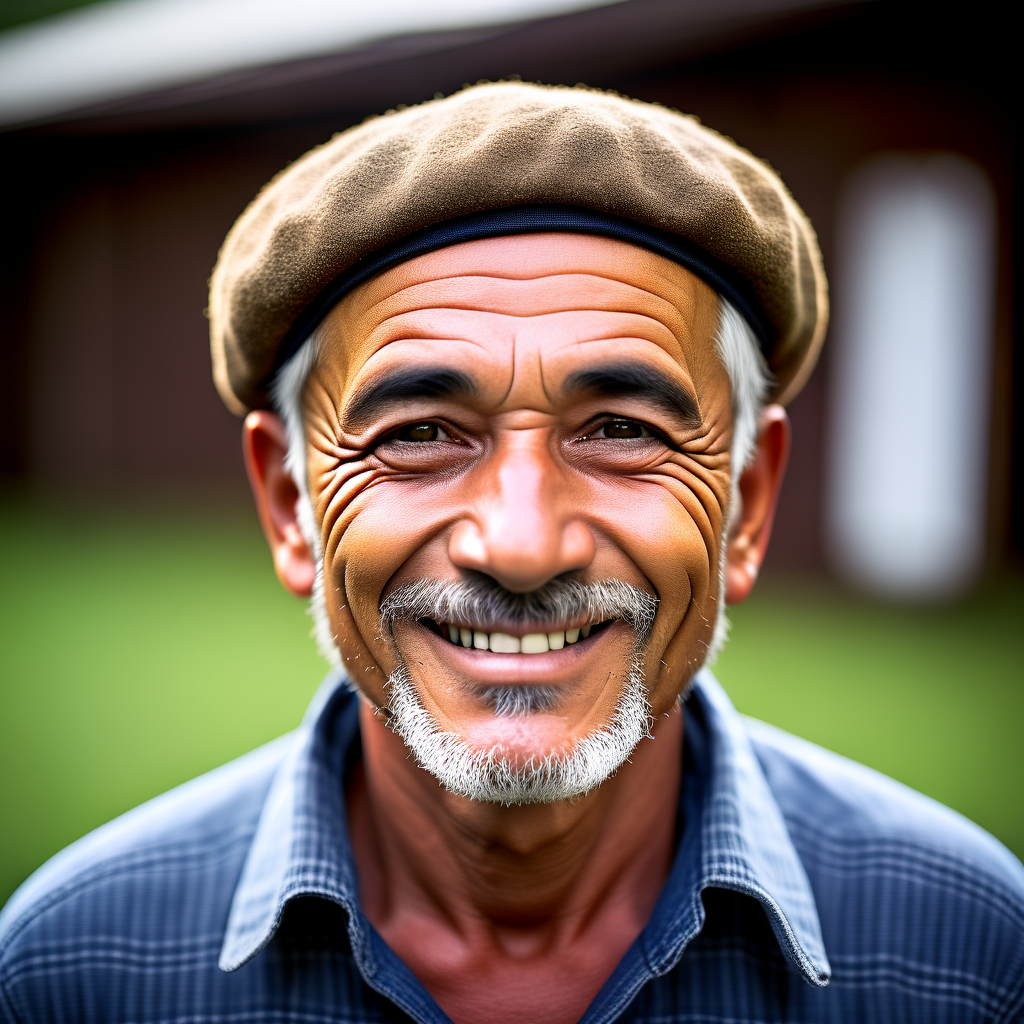}}
    \hfill
    \subfloat[SCD\label{ae}]{%
        \includegraphics[width=0.098\linewidth]{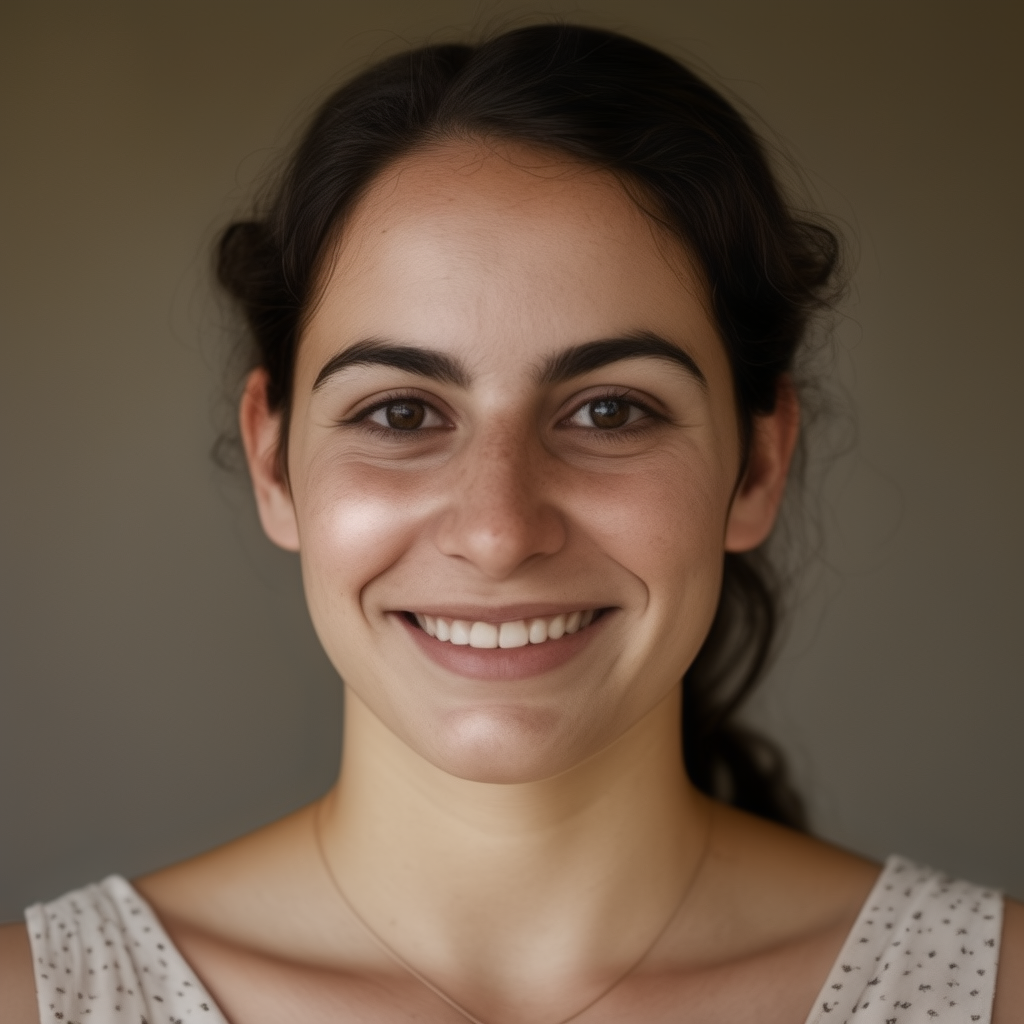}}
    \hfill
    \subfloat[SAE\label{af}]{%
        \includegraphics[width=0.098\linewidth]{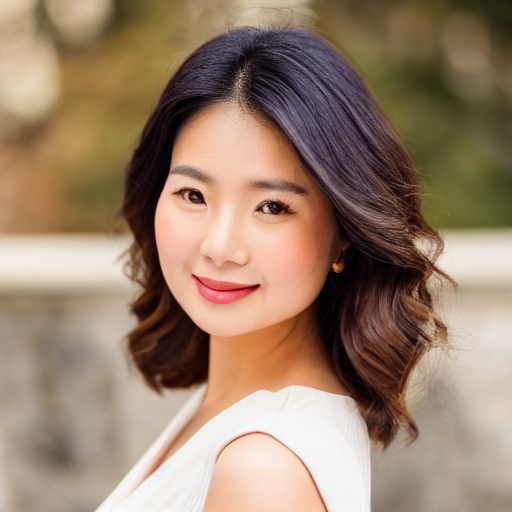}}
    \hfill
    \subfloat[SG1\label{ag}]{%
        \includegraphics[width=0.098\linewidth]{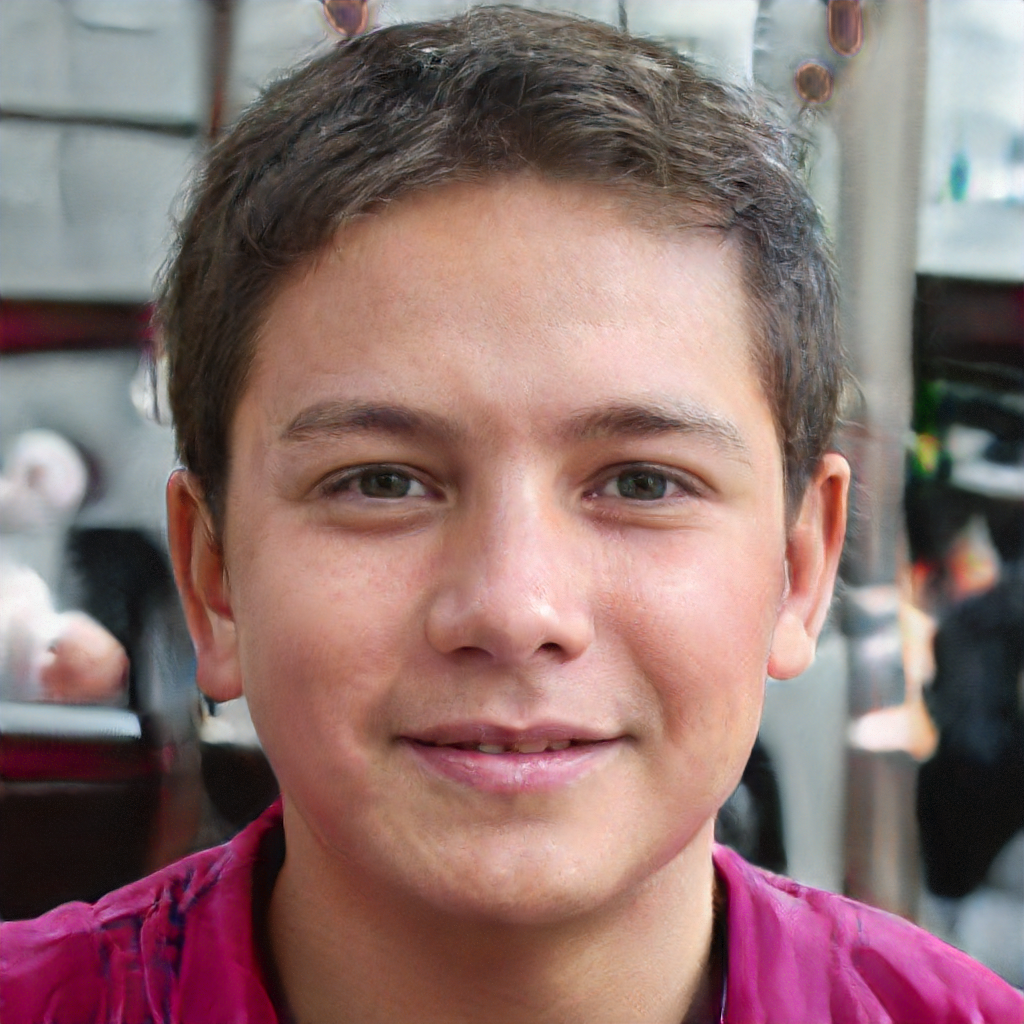}}
    \hfill
    \subfloat[SG2\label{ah}]{%
       \includegraphics[width=0.098\linewidth]{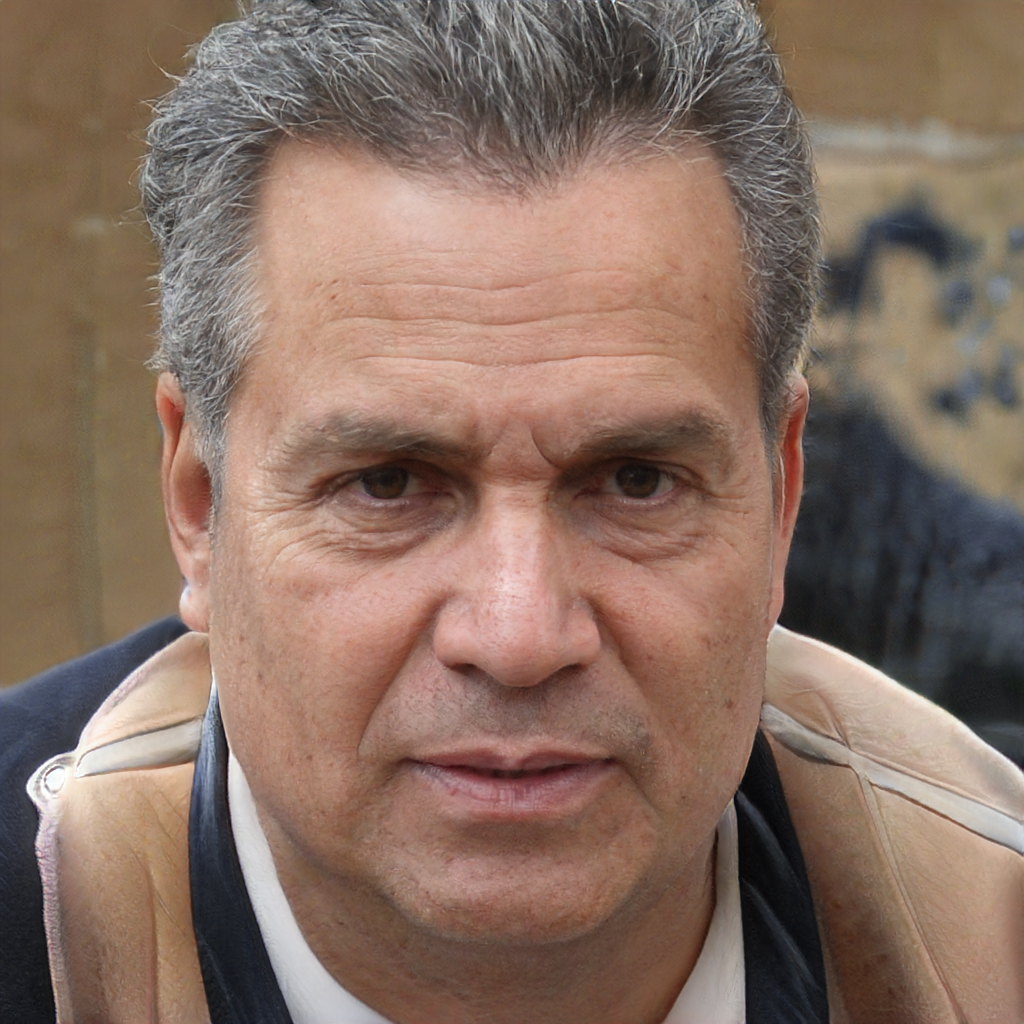}}
    \hfill
    \subfloat[SG3\label{ai}]{%
        \includegraphics[width=0.098\linewidth]{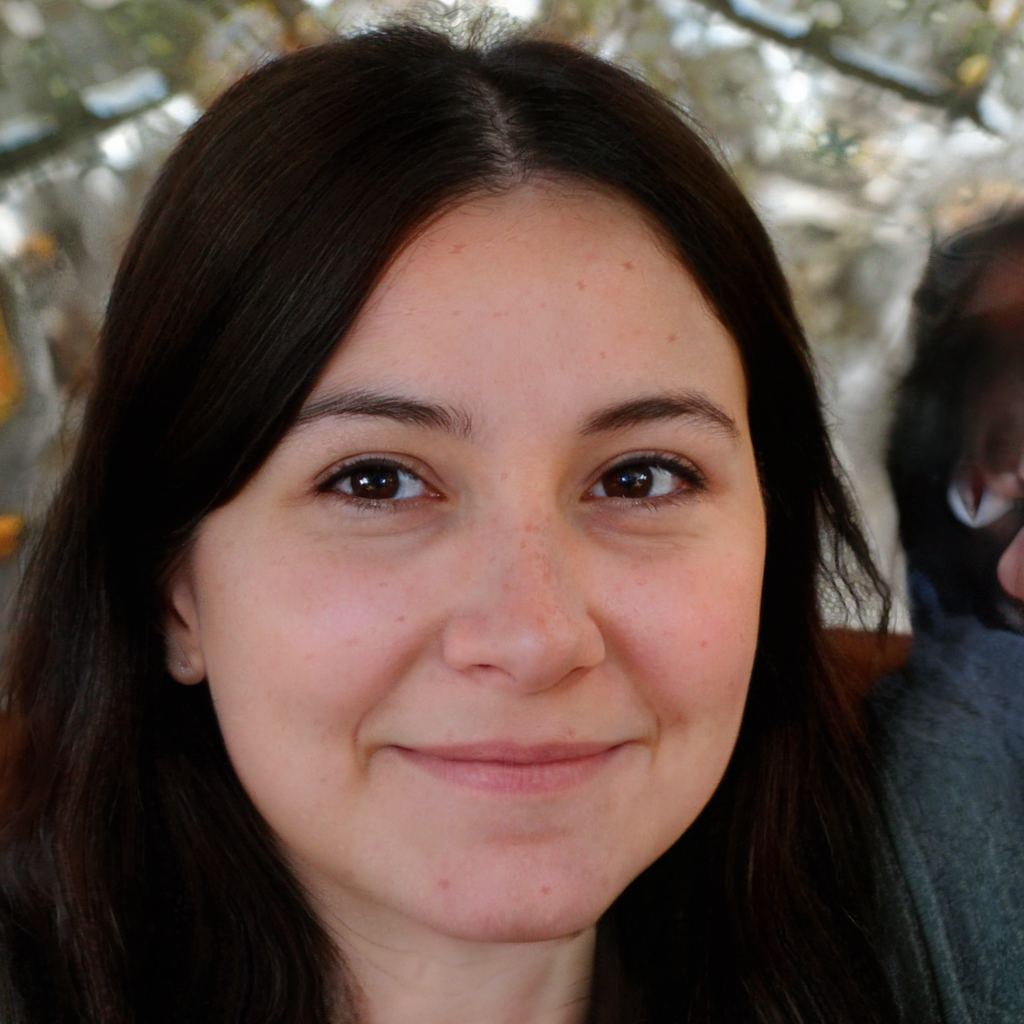}}
    \hfill
    \subfloat[THY\label{aj}]{%
        \includegraphics[width=0.098\linewidth]{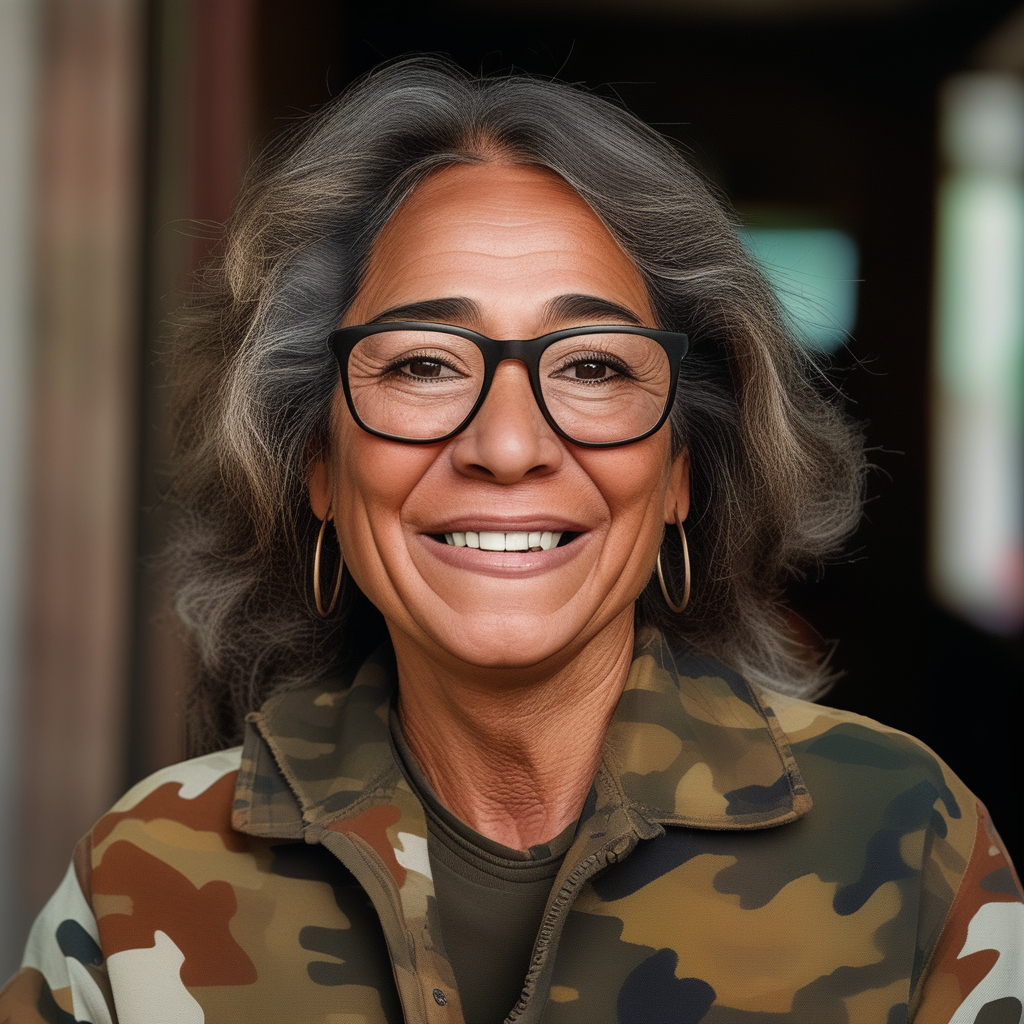}}
  \caption{Some images from the open set section of WILD (one image per generator). The generators are identified by the codes described in Section \ref{subsec:open_set}.}
  \label{fig:open_set_sample} 
  \vspace{-10pt}
\end{figure*}

To avoid biases, we managed data distributions in two different ways: subjects were described according to their gender, together with their age category ($8$ categories equally distributed over $80\%$ of the prompts) or profession ($20\%$ of the prompts). Each gender covers $50\%$ of the prompts. These characteristics were rotated to ensure a bias-free dataset along the respective axes. For profession-based descriptions, the profession was randomly drawn from a uniform distribution of 110 professions. The pronoun token followed the subject's gender. The verb tokens were simply conjugated according to the rest of the sentence. An ethnicity/nationality descriptor was randomly selected for each subject.

The $<$expression$>$ token varies depending on the sentence structure and can describe the expression (e.g., smiling, happy) or emotion shown by the subject (e.g., joy, sadness). 
The emotion/expression was randomly chosen from a distribution.

The $<$facial features$>$ token corresponds to exactly two features for every subject randomly drawn from a uniform distribution. 
Depending on the type of feature, subsequent random choices were made, including hair color, hairstyle, nose shape, eye color, or other similar details. Two instances of the same type were not allowed. To ensure correctness, some facial features were moved through the sentence from the position of the $<$facial features$>$ token to the front of the $<$subject$>$ token, for example: “Picture of a \textit{blonde} $<$subject$>$ with \textit{freckles}...".

The $<$clothing$>$ token describes one, two, or three clothing items. If multiple items appear, they must be from different categories: main clothing, headwear, accessories. Each category has a given probability of appearing, and when no category appears, the random selection is repeated until at least one does. 
The specific item to include in the description is selected from a uniform distribution for each category. Depending on the item, a following randomization can be introduced to select its pattern, color, or material (e.g., respectively: a plaid scarf, a yellow t-shirt, a silver necklace).

The $<$looking direction$>$ token introduced another random choice on where the subject is looking. This corresponded to “into the camera" with a probability of $0.8$, or to “up", “down", “left", “right" with a probability of $0.05$ each.

The $<$background$>$ token, when present, was randomized between three different description modes: color, pattern, scenery. If color was selected, a random color was specified (e.g. “... on a blue background."). If pattern was selected, the prompt describes a pattern, item or group of items that should appear behind the subject (e.g. “... with leaves in the background" or “... with a fountain in the background"). If scenery was selected, the subject is described as located in a particular place, on which the background generation will depend (e.g. “... in a village", “... on the grass", “... in the mountains").

The generated prompts were validated through a block of filters to avoid bad prompts and absurd co-occurrences of characteristics (e.g., a child with a beard). The prompts were then used to generate images. Each prompt that failed the validation or did not result in a successful generation with all the ten closed set generators was substituted with a new one, using the same sentence structure, gender, and age category (or profession), but randomizing again all the other characteristics.

Images with the same prompt are expected to have similar semantic characteristics, yet some differences are introduced by the different generative methods. These differences are due to model architecture, filters, training base (and biases), and parameters, and they are fundamental for source attribution, together with the low-level model signature. 
An example of the ten images generated with a prompt from our dataset is shown in Fig.~\ref{fig:prompt_example} (top row, a-j), where it is evident how models can give different interpretations to prompt features: the pullover, the background, and the facial features are all interpreted with different nuances, also from a semantic point of view. A similar example is shown in the bottom row (k-t) of Fig.~\ref{fig:prompt_example}. In this case, it can be observed how prompt specifications are slightly (or sometimes completely) misinterpreted by generators: the fedora specified in the prompt is not always a fedora, but maybe another type of hat, the necklace is ignored by some models. Moreover, some details are added or extended to cover what is not specified in the prompt: the rest of the clothing might take the same color specified for the fedora or can have a different pattern, like those produced by Leonardo AI and Stable Diffusion 3.5. When a background is not specified, instead, a plausible one might be inferred from other details.

\subsection{Open Set}
\label{subsec:open_set}
To simulate a real-world case scenario where detection models need to attribute generators in the wild, WILD features an open set section with $10,000$ synthetic head and shoulder images obtained from another group of $10$ generators. In this case, we set no constraints on the prompts and data distributions. Moreover, to create the open set, we did not use exclusively text-to-image models. 
In the following, we introduce the list of generative methods used for the open set. Also in this case, we use a three-character code to identify them in tables and figures: DALL$\cdot$E\, 1 (DE1) \cite{ramesh2021zeroshot}, Deep AI (DPA) \cite{deepai}, Hotpot AI (HPA) \cite{hotpotai}, NVIDIA Sana PAG (NVS) \cite{xie2024sana},
Stable Cascade (SCD) \cite{pernias2023wuerstchen}, Stable Diffusion Attend\&Excite (SAE) \cite{sd_ae}, StyleGAN (SG1) \cite{karras2019stylebasedgeneratorarchitecturegenerative}, StyleGAN2 (SG2) \cite{karras2020analyzingimprovingimagequality}, StyleGAN3 (SG3) \cite{karras2021alias}, Tencent Hunyuan (THY) \cite{li2024hunyuandit}.
Some examples of images included in the open set are displayed in Fig.~\ref{fig:open_set_sample}, showing one image for each generator of the open set.

\subsection{Post-Processing}
\label{subsec:post_processing}
For post-processing, we applied various image enhancement and processing techniques to half of the images of the dataset (both closed set and open set). This simulates real-world conditions where manipulated images may undergo transformations that obscure the detection cues. Each image underwent one to three randomly selected post-processing operations chosen from a set of ten transformations: JPEG/WebP compression with a quality factor in the range \([50,100]\); random central cropping with up to 10\% border loss; resizing with a scaling factor in the range \([0.4,2]\); rotation by \(\pm 0.5^\circ\) with aspect-preserving cropping; contrast/brightness adjustment with factors in the range \([0.8,1.2]\);  Gaussian blur with a radius in the range \([0.5,1.5]\);  grayscale conversion; super-resolution using pre-trained models from NinaSR \cite{ninasr}, CARN \cite{ahn2018fast}, EDSR \cite{lim2017enhanced}, RDN \cite{zhang2018residual}, or RCAN \cite{zhang2018image}; JPEG AI compression~\cite{ascenso2023jpeg} with variable Bit Per Pixel (BPP) compression rate in the set $[0.5, 0.75, 1, 1.25, 1.50]$. 
Post-processed images were organized by their transformation depth ($1$ to $3$ steps), enabling controlled testing of detector robustness against increasingly complex modification chains. The post-processed images were not included in the training set and were used exclusively during testing.

\subsection{Dataset Split}
\label{subsec:split}
The closed set is released with a predefined split to facilitate cross-comparisons. The split was operated at the prompt level, to avoid images with the same prompt ending up in different sets, and to ensure a perfect class balance between closed set generators. The training, validation, and test sets contain respectively $5,000$ images ($500$ prompts), $2,000$ images ($200$ prompts), and $3,000$ images ($300$ prompts).
The splitting procedure took into account the prompt blocks described in Section~\ref{subsec:prompts} and respected the global data distributions, keeping biases at a minimum level. Each set has the same percentage of prompts for every gender ($50\%$) and for every age category and the same percentage of profession-based prompts. Moreover, the five prompt structures are distributed equally among the sets. In contrast, the distributions of ethnicity/nationality, facial features, clothing, background, and expression/emotion of the characters were randomized during the prompt building procedure. As a consequence, we did not enforce their uniform distribution among sets. All the test set images have post-processed versions. The open set instead was not split, as it is not intended to be used for training.
\section{Baselines and Results}
\label{sec:baselines}

In the following, we assess the performance of some common attribution baselines on our proposed WILD dataset. 
We start by describing the baselines selected for the benchmark, then we show their results on closed set attribution and open set attribution.


\subsection{Baseline methods}
\label{subsec:methods}

\subsubsection{Clip Feature Classifiers}
\label{subsubsec:clip_plus}
CLIP (Contrastive Language–Image Pre-training)\cite{radford2021learning} is used as a feature extractor in a wide range of tasks. Being pre-trained on large amounts of data, it allows to train classifiers directly on its features, even when few data are available. 
We used CLIP to extract image features without specifying the prompt. The features were then passed to a small classification head, which was either a Multi-Layer Perceptron (MLP) or a Support Vector Machine (SVM). We used CLIP Large with input size $336$. The hyperparameters of the two models were tuned on the validation set. For CLIP+MLP, we used an MLP with two hidden layers of $512$ and $256$ units, respectively, both with ReLU activation and softmax output layer. The MLP was trained with Adam \cite{Adam}, an initial learning rate of $10^{-3}$ and $l_{2_\alpha}$ of $10^{-4}$, for a maximum of $1,000$ epochs and the validation loss as early stopping criterion, with patience $10$ (restoring the best weights). The SVM instead used a Radial Basis Function (RBF) kernel, with C equal to $1.0$, a tolerance of $10^{-3}$, degree $3$, and no maximum number of iterations.

\subsubsection{DE-FAKE}
\label{subsubsec:defake}
In \cite{sha2023fake}, Sha et al. introduced a novel framework for synthetic image source attribution: a hybrid classifier for robust detection and attribution across different generative techniques. The model exploits multimodal features extracted from both the input image and its prompt description through CLIP’s image and text encoder \cite{radford2021learning}. When prompt descriptions are not available, they are estimated using Blip2 \cite{li2022blip}. 
During our experiments, DE-FAKE \cite{sha2023fake} was trained and tested following the procedures and using the hyper-parameters proposed in the original publication. 

\subsubsection{Standard \glspl{cnn}}
\label{subsubsec:cnn}
As additional baselines, we considered classic \glspl{cnn} which have been successfully used in many forensic tasks. Specifically, we selected the EfficientNetB4~\cite{tan2019efficientnet}, XceptionNet~\cite{chollet2017xception}, and ResNet50~\cite{he2016deep} architectures.
We trained them following a similar approach to that proposed in~\cite{mandelli2024synthetic}: each \gls{cnn} works at patch level, considering square RGB patches of $96 \times 96$ pixels as input and providing a single score per patch. 
At deployment stage, given an image, we randomly extracted $50$ patches and computed the final score for the image as the arithmetic mean of the patch scores.
We adopted this approach as it demonstrated strong robustness to compression and resizing operations, thanks to an extensive set of augmentations incorporated into the training process~\cite{mandelli2024synthetic}.
The ability to extract small patches from the query image reduces the dependency on its semantic content, allowing for a sharper focus on the synthetic generation artifacts.

\subsubsection{Vision Transformer Classifier}
\label{subsubsec:vit}
Vision Transformers (ViT)~\cite{dosovitskiy2020image} are a valid alternative to CNNs for synthetic image detection. ViTs leverage self-attention to capture long-range dependencies.
Along with the other ViT-based methods, which exploit the ViT for feature extraction, we also employed a ViT directly as a classifier. For the sake of clarity, we call it Vision Transformer Classifier (VTC) in the following.
We employed the ViT-Base model, pre-trained on ImageNet~\cite{deng2009imagenet} and ImageNet-21k~\cite{ridnik2021imagenet21kpretrainingmasses} and fine-tuned it for synthetic image attribution on our closed set. Images were split into $16 \times 16$ patches, embedded, and processed by the transformer encoder.
We aggregated patch-level predictions to obtain the final classification score, enhancing resilience against localized distortions.

\subsection{Closed Set Attribution}
\label{subsec:csa}
We evaluated the models' performance in terms of balanced accuracy for closed-set attribution. This is the simplest task, as the evaluation is limited to the ten generators encountered during training. 
The results in Table \ref{tab:closed_set_attribution} show that CNN-based methods, known to learn low-level features, significantly outperform other baselines on plain images. 
In contrast, their performance drops sharply when evaluated on post-processed images, reporting the worst results among all the investigated methodologies. 
The VTC method, which relies on higher-level features, proves to be more robust to post-processing. 
These findings align with recent observations in \cite{cozzolino2024raising}, which indicate that ViT-based solutions are more suitable for use in the wild.

\begin{table}[t]
\caption{Balanced accuracy on the closed set attribution task for all the baselines, on the plain and post-processed (1,2,3 steps) image versions.}
\centering
\resizebox{.8\columnwidth}{!}{
\begin{tabular}{@{}lcccc@{}}
\toprule
Baseline & Plain & 1 Step & 2 Steps & 3 Steps \\
\midrule
CLIP+MLP & 96.67\% & 87.13\% & 80.13\% & 72.40\% \\
CLIP+SVM & 95.30\% & 86.50\% & 78.93\% & 72.63\% \\
DE-FAKE & 94.00\% & 87.00\% & 83.00\% & 78.00\% \\
EfficientNetB4 & \textbf{100.0\%} & 84.43\% & 73.03\% & 59.93\% \\
XceptionNet & \textbf{100.0\%} & 84.00\% & 72.67\% & 58.70\% \\
ResNet50 & 99.90\% & 84.87\% & 73.53\% & 60.63\% \\
VTC & 95.80\% & \textbf{93.40\%} & \textbf{91.50\%} & \textbf{88.67\%} \\
\bottomrule
\end{tabular}
}
\label{tab:closed_set_attribution}
\end{table}

\subsection{Open Set Results}
\label{subsec:osa}
For the open set attribution task, we adopted two different approaches. 
We started by modeling open set attribution as a binary classification problem, i.e., separating closed set samples from open set ones. Then, we tested the baselines on multi-class attribution, considering, together with the open set samples, all the closed set classes. 

\subsubsection{Binary classification}
We used the scores provided by the attribution models over the samples coming from the test set of the closed set and from the open set. 
We selected the maximum softmax scores achieved over all the ten closed set classes.
Then, we exploited the distribution of these scores to classify the test samples as coming from the ``closed set'' class (positive) or from the ``open set'' one (negative). The ``open set'' class is reasonably associated with low softmax probabilities.
To evaluate the models' performance, we measured the binary ROC-AUC, the Equal Error Rate (EER), and the True Positive Rate evaluated at False Positive Rate equal to $5\%$ (TPR@FPR=0.05). 
Table \ref{tab:open_set_binary} reports the results on non-post-processed test set samples.
Similarly to what has been observed for the closed-set, these results highlight the strong capability of the three CNN-based methods to discriminate closed-set samples from open-set ones, with EfficientNetB4 achieving the best results on all the metrics.

\begin{table}[t]
\caption{Open-set vs closed-set classification results (without post-processing).}
\centering
\resizebox{.8\columnwidth}{!}{
\begin{tabular}{@{}lccc@{}}
\toprule
Baseline & ROC-AUC & EER & TPR@FPR=0.05 \\
\midrule
CLIP+MLP& 0.876& 0.202& 0.373 \\
CLIP+SVM& 0.872& 0.210& 0.385 \\
DE-FAKE& 0.836& 0.244& 0.485\\
EfficientNetB4& \textbf{0.997}& \textbf{0.028}& \textbf{0.984}\\
XceptionNet& 0.992& 0.046& 0.961\\
ResNet50& 0.985& 0.061& 0.923\\
VTC& 0.890& 0.181& 0.374\\
\bottomrule
\end{tabular}
}
\vspace{-10pt}
\label{tab:open_set_binary}
\end{table}

In Table~\ref{tab:open_set_binary_pp}, we compare the three best state-of-the-art methodologies (considering only a single approach among the CNN-based ones) over post-processed images. 
As in the closed-set case, VTC tends to be the most robust approach, even if the TPR@FPR=0.05 is still quite low and sometimes outperformed by CNN-based solutions.

\begin{table}[t]
\caption{Binary closed vs open set classification results for CLIP+MLP, EfficientNetB4, and VTC on post-processed images.}
\centering
\resizebox{.9\columnwidth}{!}{
\begin{tabular}{@{}lccc@{}}
\toprule
Metric(Steps) & CLIP+MLP & EfficientNetB4 & VTC \\
\midrule
ROC-AUC (1)& 0.781 & \textbf{0.872} & 0.857 \\
ROC-AUC (2)& 0.720 & 0.763 & \textbf{0.829} \\
ROC-AUC (3)& 0.662 & 0.659 & \textbf{0.801} \\
\midrule
EER (1)& 0.284 & \textbf{0.202} & 0.215 \\
EER (2)& 0.341 & 0.308 & \textbf{0.240} \\
EER (3)& 0.384 & 0.389 & \textbf{0.271} \\
\midrule
TPR@FPR.05 (1)& 0.180 & \textbf{0.560} & 0.256 \\
TPR@FPR.05 (2)& 0.122 & \textbf{0.305} & 0.197 \\
TPR@FPR.05 (3)& 0.089 & 0.163 & \textbf{0.169} \\
\bottomrule
\end{tabular}
}
\vspace{-10pt}
\label{tab:open_set_binary_pp}
\end{table}

\begin{figure*} 
    \centering
     \subfloat[Plain\label{3a}]{%
       \includegraphics[width=0.24\linewidth]{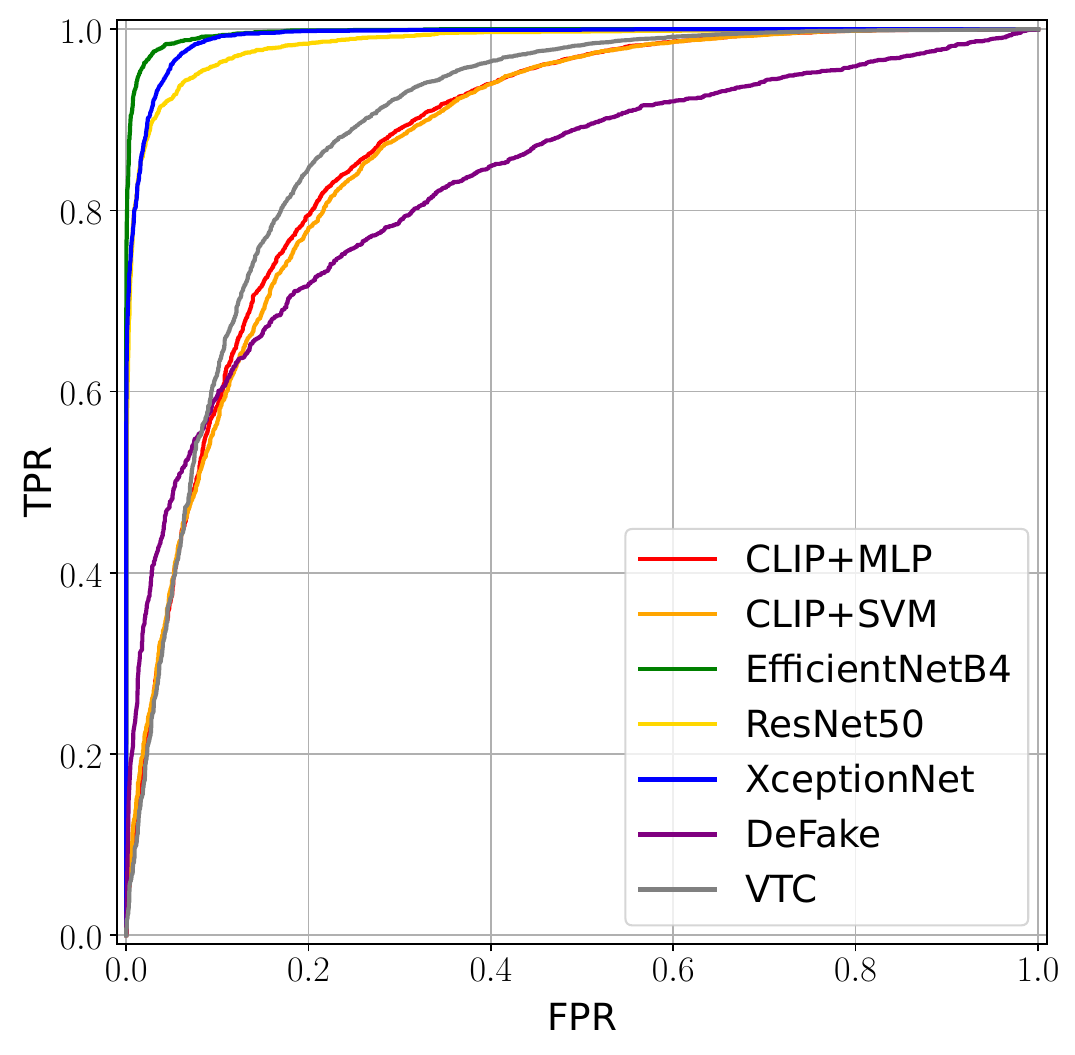}}
    \hfill
    \subfloat[1 step\label{3b}]{%
        \includegraphics[width=0.24\linewidth]{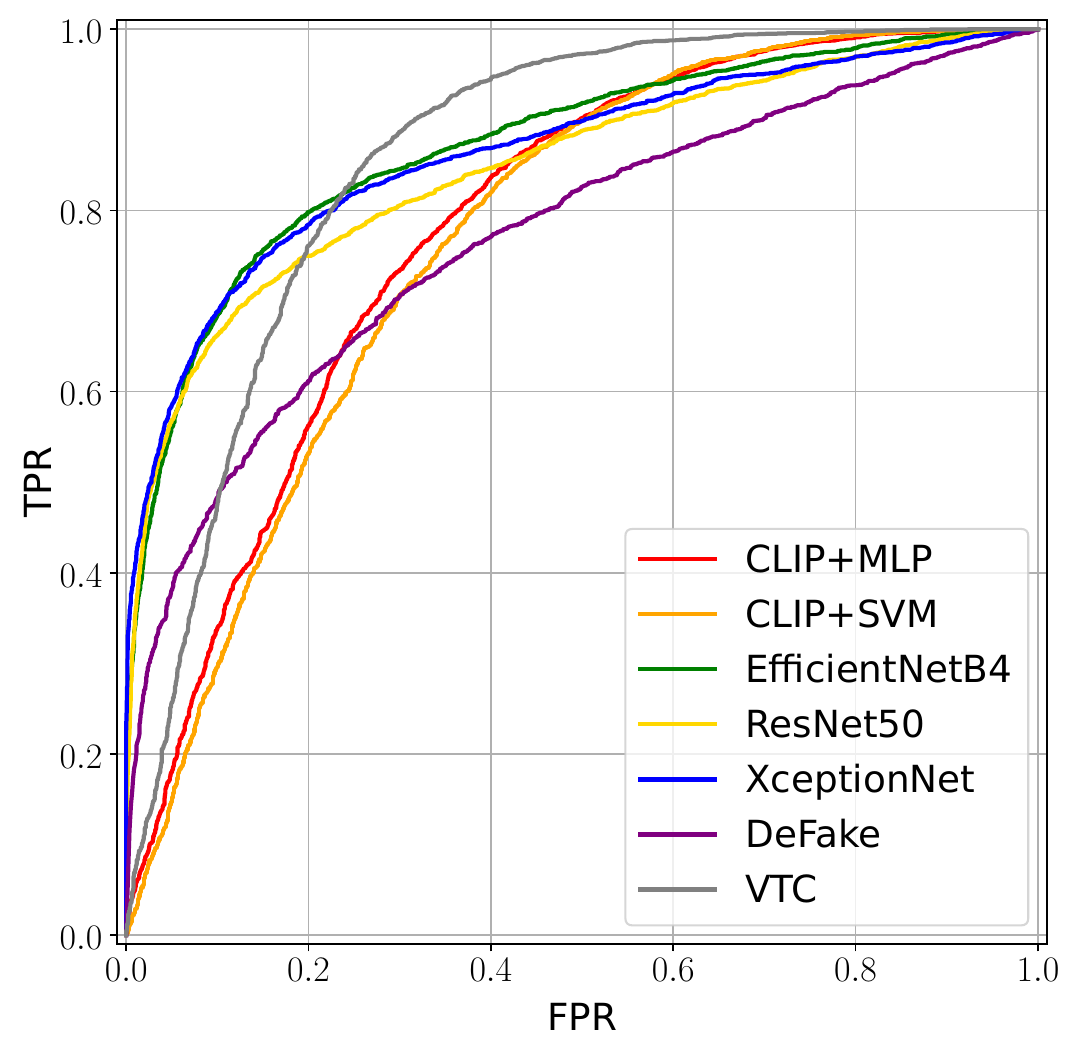}}
    \hfill
    \subfloat[2 steps\label{3c}]{%
       \includegraphics[width=0.24\linewidth]{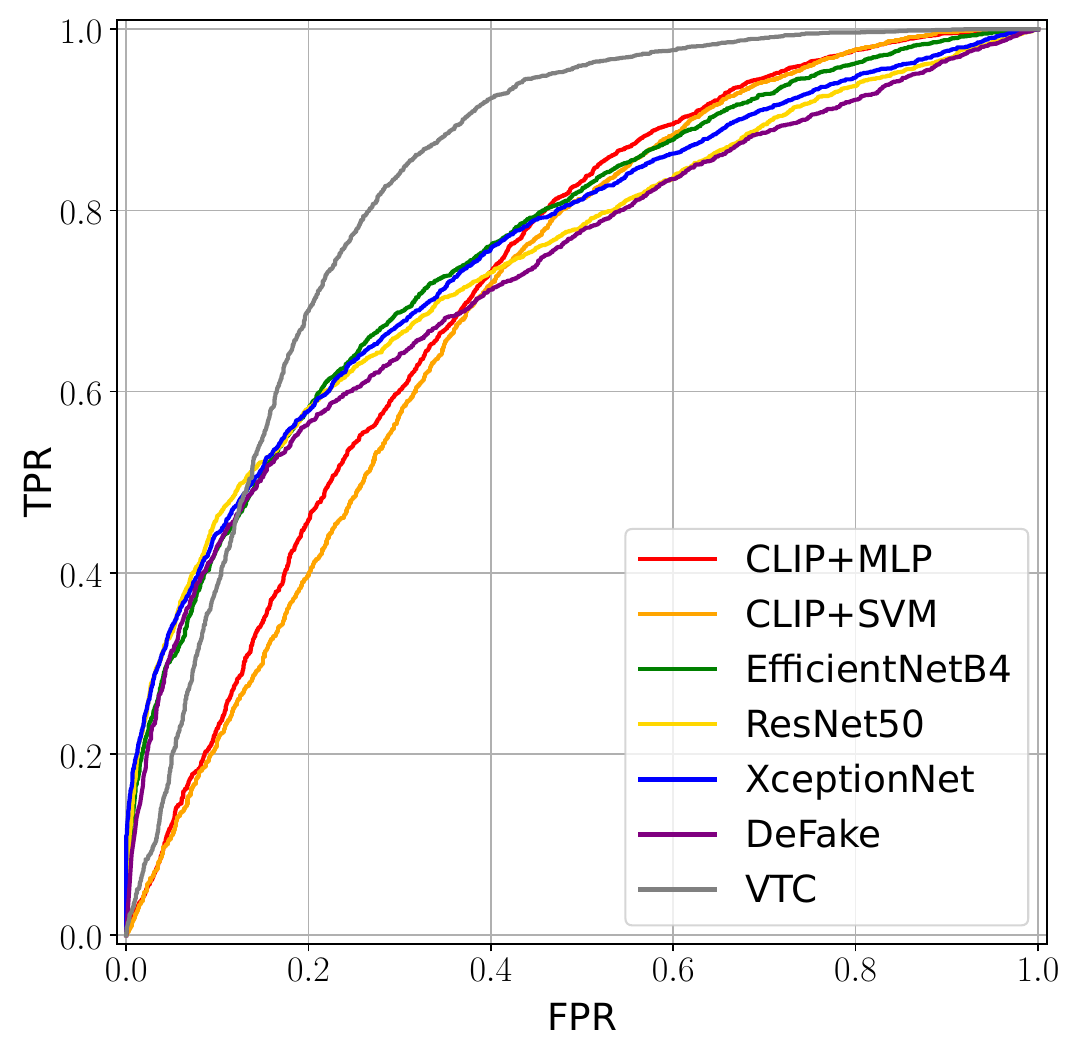}}
    \hfill
    \subfloat[3 steps\label{3d}]{%
        \includegraphics[width=0.24\linewidth]{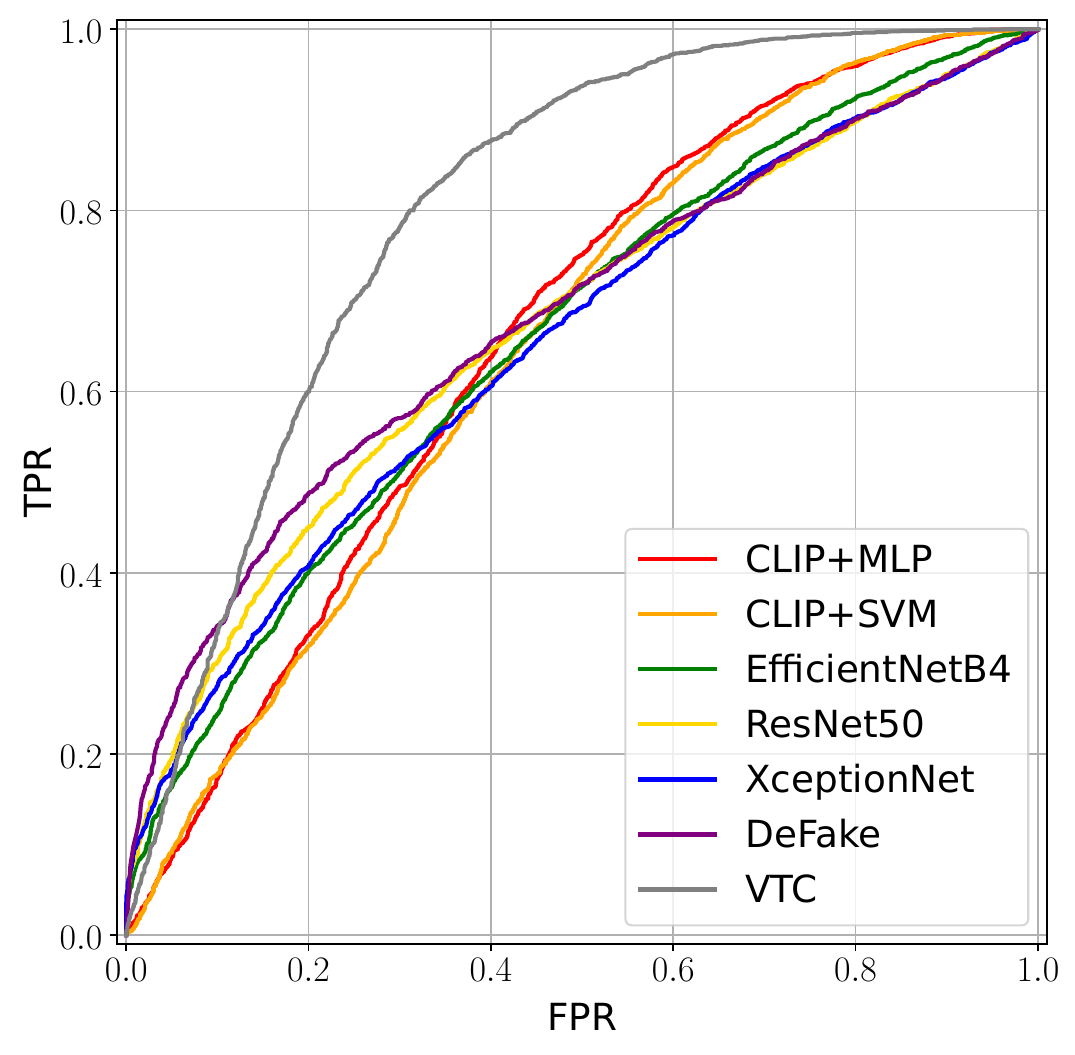}}
  \caption{ROC Curves of baselines in the binary classification task between open set (positive class) and closed set (negative task), with different levels of image post-processing: (a) plain images, (b) 1 post-processing step, (c) 2 post-processing steps, (d) 3 post-processing steps.}
  \label{fig:binary_roc} 
\end{figure*}

Finally, Fig.~\ref{fig:binary_roc} presents the ROC curves obtained for both plain and post-processed images, confirming our previous findings. CNN-based methods significantly outperform other baselines on plain images, with EfficientNetB4 demonstrating the best ROC curve, closely followed by ResNet50 and XceptionNet. However, their advantage diminishes on post-processed images. With three post-processing steps, VTC emerges as the best-performing model, while the other methods exhibit comparable performance. This again highlights that high-level features like those extracted by ViT-based methods contribute to greater model robustness.


\subsubsection{Multi-class classification results}
\label{multi_class_open_set}
\begin{figure*}[t]
    \centering
     \subfloat[CLIP+MLP\label{4a}]{%
       \includegraphics[width=0.25\linewidth]{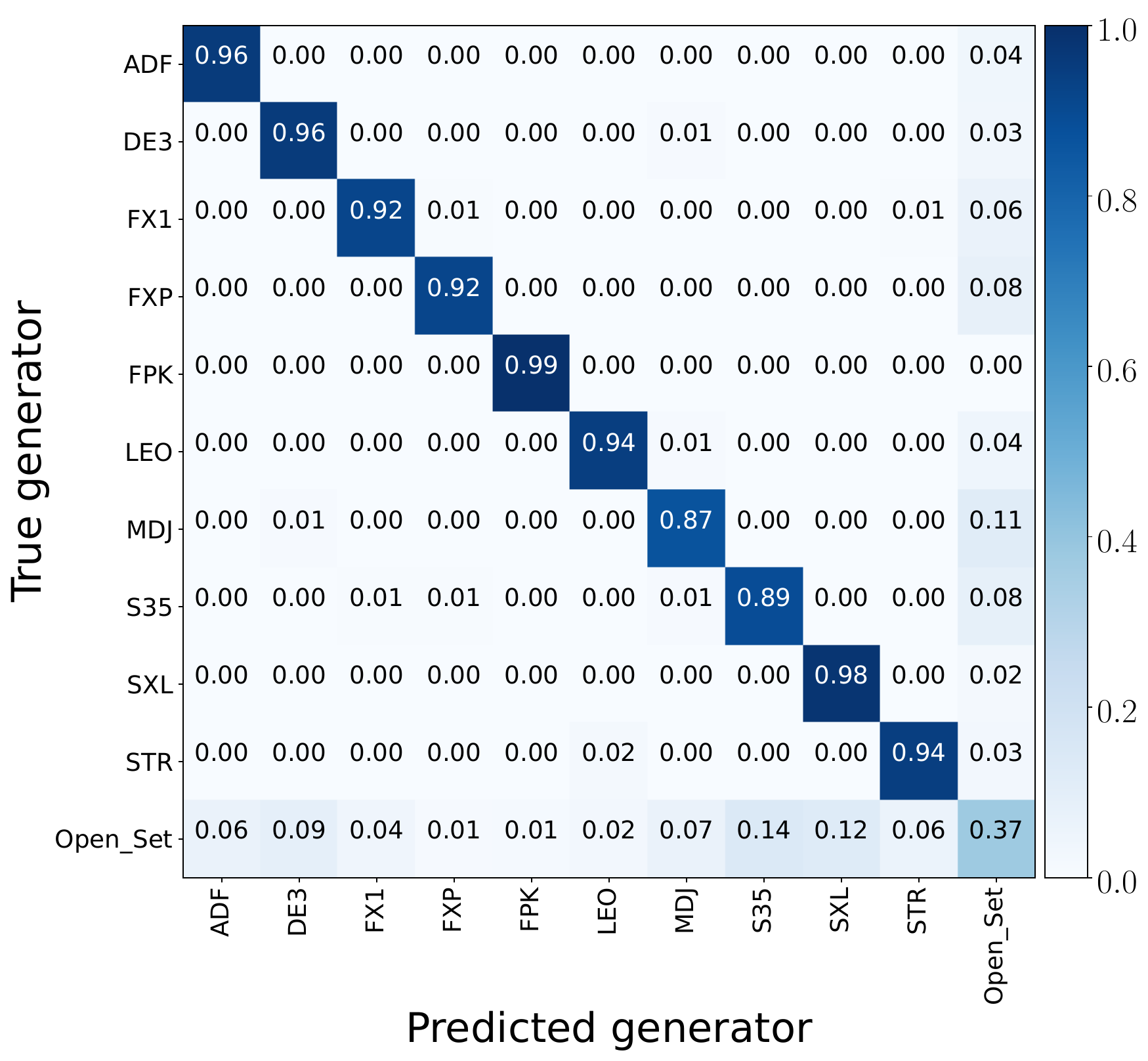}}
    \hfill
    \subfloat[DE-FAKE\label{4b}]{%
       \includegraphics[width=0.25\linewidth]{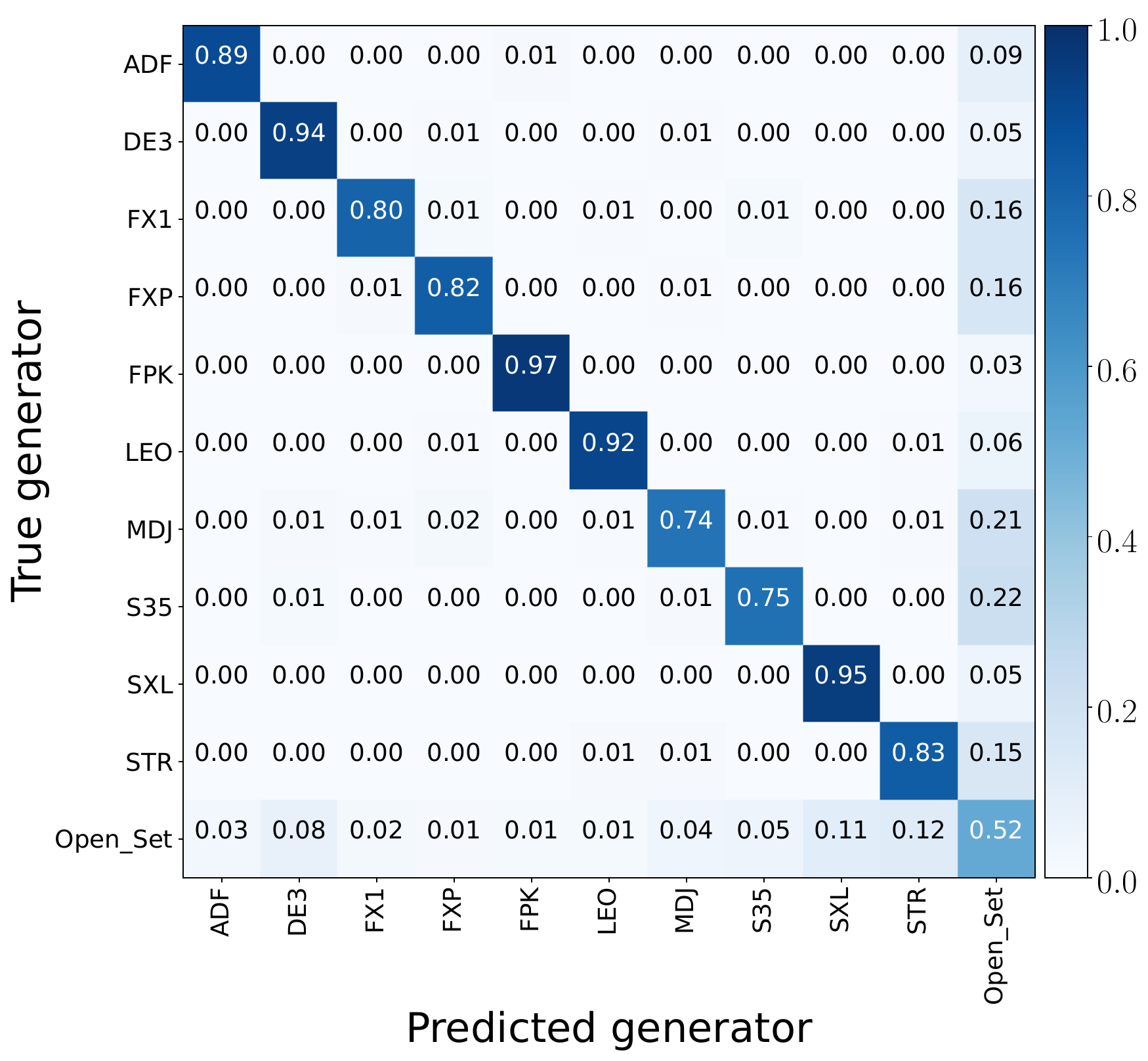}}
    \hfill
    \subfloat[XceptionNet\label{4c}]{%
        \includegraphics[width=0.25\linewidth]{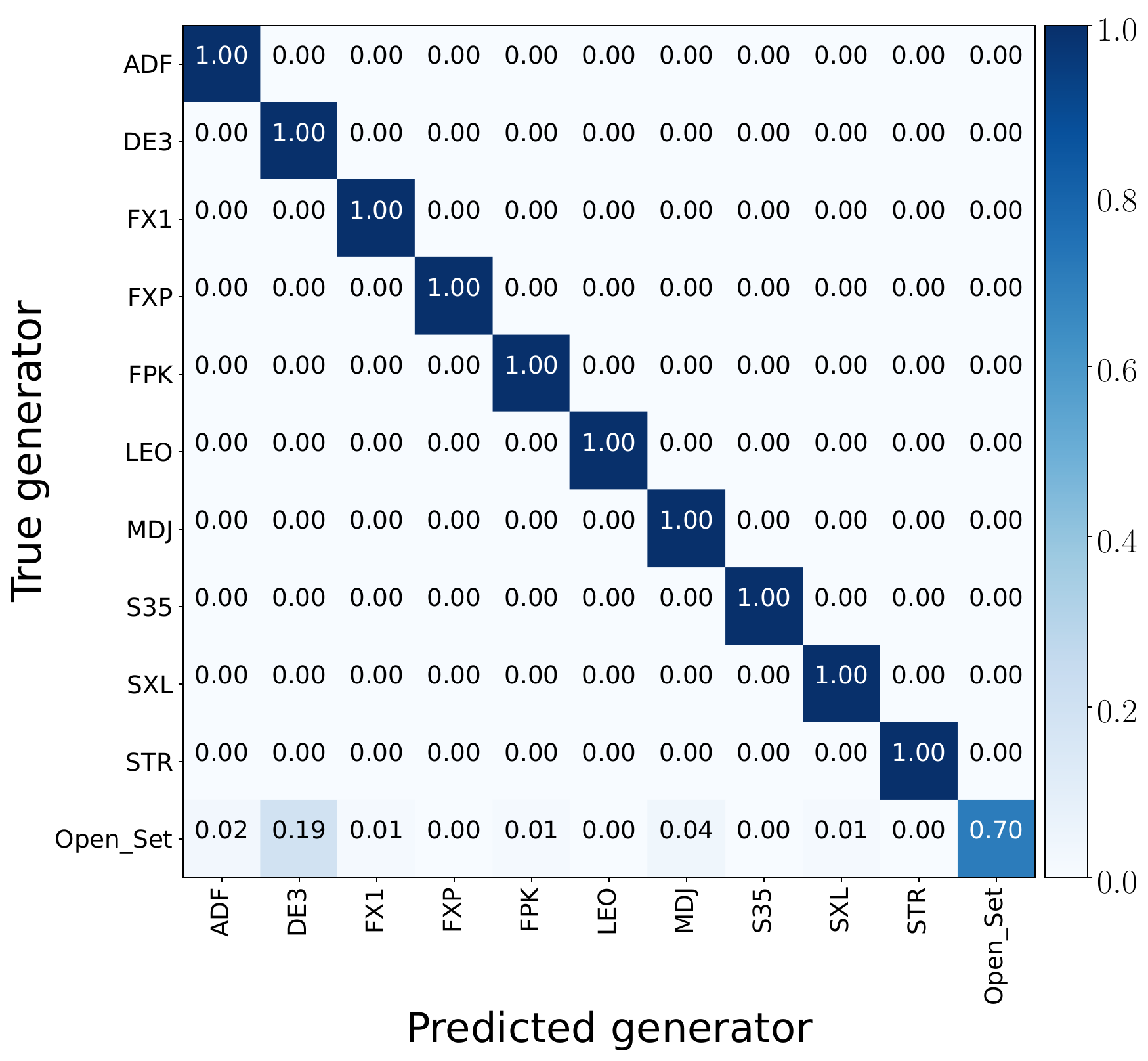}}
    \hfill
    \subfloat[VTC\label{4d}]{%
       \includegraphics[width=0.25\linewidth]{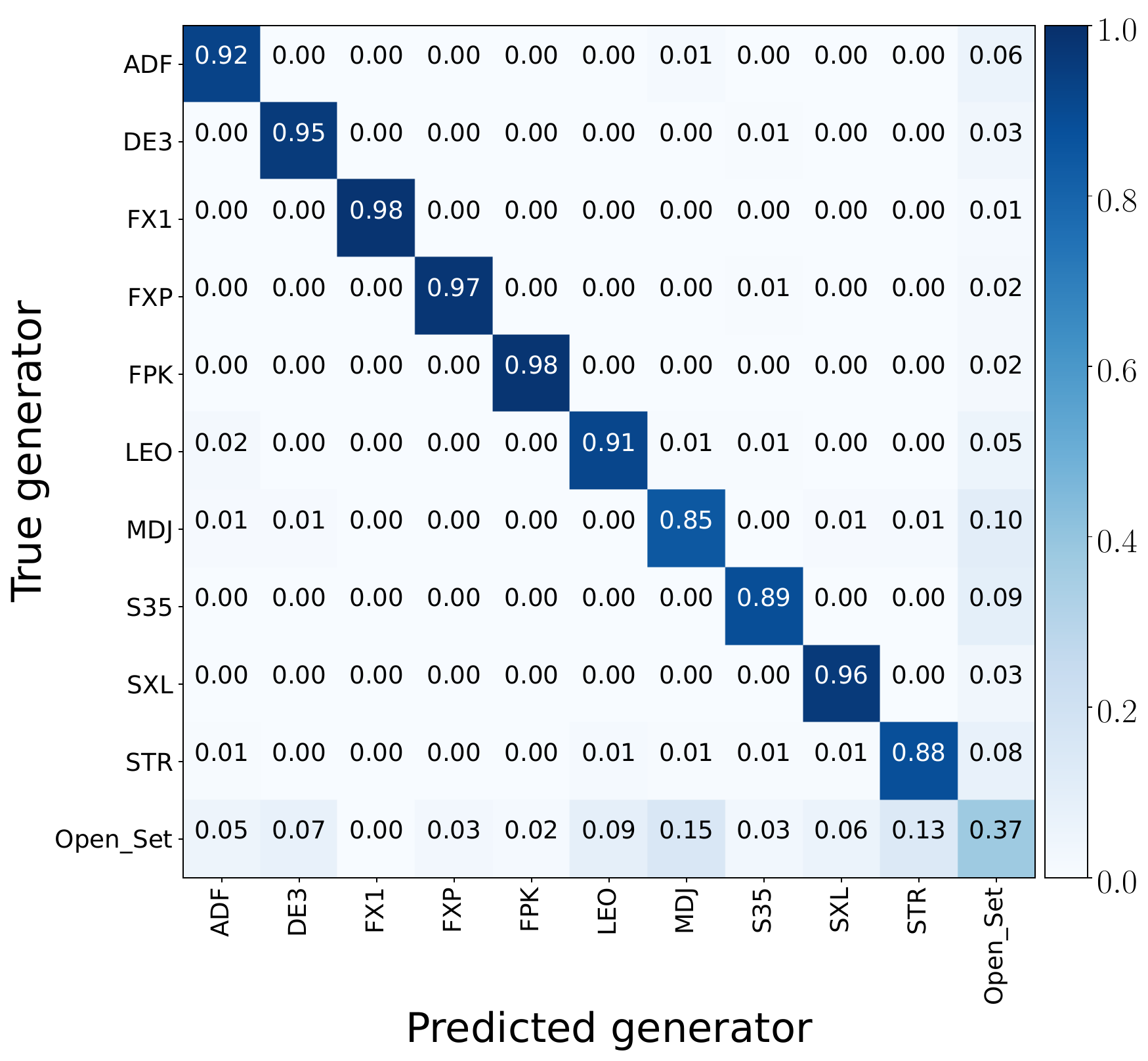}}
  \caption{Confusion matrices obtained by CLIP+MLP (a), DE-FAKE (b), XceptionNet (c), and VTC (d) on the open set attribution with rejection threshold, without post-processing. The generator-classes are represented by the three-letter codes defined in Section \ref{subsec:generators}.}
  \label{fig:cm_all} 
  \vspace{-10pt}
\end{figure*}

In this setup, we evaluated the performance of the detectors in attributing both the closed set and open set samples. To do this, we followed a two-step approach. First, we fixed a rejection threshold on the softmax probabilities. Then, if no class probability was larger than the threshold, the example was rejected, and classified as unknown or “open set'', otherwise we attributed it to the class with maximum probability.
We fixed the rejection threshold on the validation set samples. We recall that the validation set contains only non-processed, closed set samples. 
We then extracted the maximum softmax scores and 
set the threshold to obtain a rate of incorrect detections (i.e., samples detected as being ``open set'') equal to $5\%$.
At test time, we applied the rejection threshold to samples of the closed set (test set) and the open set.
We counted how many samples were correctly rejected but also how much the rejection impacted the attribution capabilities on known classes. 
Fig.~\ref{fig:cm_all} shows the obtained confusion matrices.  
Notice that the ``open set'' has been added as an additional class.  
For the sake of readability, we do not report the results of all the baselines, but only the most interesting ones. The Correct Classification Rate (CCR) of every closed set class can be read in the diagonal cells, as well as the Correct Rejection Rate (CRR) of the open set (last diagonal cell). 
In this task, XceptionNet neatly outperforms all the other baselines: having a perfect CCR of $1.0$ on all closed set classes (which is true also for EfficientNetB4 and ResNet50), it also shows reasonably good rejection performance with CRR equal to $0.7$.

Concerning robustness against post-processing operations, we report in Table~\ref{tab:fixed_threshold} the average CCR and CRR of the three best methodologies. 
Concerning the closed set performance, it is interesting to notice that XceptionNet degrades rapidly its CCR on every class, confirming our previous considerations on the poor robustness (regarding the closed set attribution capabilities) of CNN-based baselines to editing operations. 
A similar behavior, even if less steep, is shown by DE-FAKE, whose CCR decreases as the number of editing steps increases. 
The best closed set attribution method appears again to be VTC, with a relatively large CCR even in the presence of post-processing.
With regard to the \textit{open set}, XceptionNet proves to be the best methodology overall, even if all the investigated techniques show a drop in performance when the images go through multiple processing steps. 

\begin{table}[t]
\caption{CRR and average CCR of XceptionNet, DE-FAKE, and VTC in open set attribution by multi-class classification with rejection.}
\centering
\resizebox{.8\columnwidth}{!}{
\begin{tabular}{@{}lccc@{}}
\toprule
Metric(Step) & XceptionNet & DE-FAKE & ViT \\
\midrule
Avg. CCR (Plain) & 1.00 & 0.86 & 0.93 \\
Avg. CCR (1 Step)& 0.79 & 0.77 & 0.91 \\
Avg. CCR (2 Steps)& 0.61 & 0.72 & 0.89 \\
Avg. CCR (3 Steps)& 0.43 & 0.65 & 0.87 \\
\midrule
CRR (Plain) & 0.70 & 0.52 & 0.37 \\
CRR (1 Step)& 0.67 & 0.52 & 0.26 \\
CRR (2 Steps)& 0.66 & 0.49 & 0.20 \\
CRR (3 Steps)& 0.64 & 0.49 & 0.17 \\
\bottomrule
\end{tabular}
\vspace{-10pt}
}
\label{tab:fixed_threshold}
\end{table}

\section{Conclusions}
\label{sec:conclusion}

We introduced WILD: a new in-the-wild Image Linkage Dataset for synthetic image attribution. The dataset consists of a closed set and an open set, composed of $10,000$ synthetic images each, enabling the evaluation of synthetic source attribution models in both closed-set and open-set scenarios. Furthermore, $5,000$ images from each set were post-processed with one, two, and three operators, simulating real-world transformations that images may undergo. 
To assess WILD, we evaluated seven common source attribution methodologies, providing a benchmark for future research on this dataset.  
Our findings show that, while synthetic image source attribution is a relatively easy task in a closed set without post-processing, 
the problem is far from being solved in real-world scenarios. 
In-the-wild conditions, where unknown generators and post-processing steps introduce significant variability, led to substantial performance degradation.
We believe that WILD represents a challenging dataset for benchmarking the next generation of image source attribution models: It allows to develop models that can work in the wild and tackle real-world attribution tasks. Moreover, as underlined by robustness evaluations, model robustness can be tested and developed on the post-processed images, which proved to be a challenge for most baselines.
WILD also opens rooms for interesting future directions: for example, it allows to investigate how prompt semantics affect different generators, but also to conduct explainability studies on source attribution methods.

\vspace{-4pt}
\section*{Acknowledgement}
This work was partially supported by project SERICS (PE00000014) under the MUR National Recovery and Resilience Plan funded by the European Union - NextGenerationEU, and by project FOSTERER, funded by MUR within the PRIN 2022 program under contract 202289RHHP.
\vspace{-4pt}
\bibliographystyle{IEEEtran}
\bibliography{bibliography}

\end{document}